\documentclass{article}

\usepackage{arxiv}
\usepackage{lineno}
\usepackage[utf8]{inputenc} 
\usepackage[T1]{fontenc}    
\usepackage{hyperref}       
\usepackage{url}            
\usepackage{booktabs}       
\usepackage{amsfonts}       
\usepackage{nicefrac}       
\usepackage{microtype}      
\usepackage{graphicx}
\usepackage[numbers]{natbib}
\usepackage{subfigure}
\usepackage{makecell}
\usepackage{doi}

\title{Dynamic Range of SiPMs with High Pixel Densities}

\date{July 25, 2024}	

\author{
  Zhiyu Zhao \\
  Tsung-Dao Lee Institute, Shanghai Jiao Tong University\\
  1 Lisuo Road, Shanghai, 201210, China \\
  Institute of Nuclear and Particle Physics, School of Physics and Astronomy\\
  800 Dongchuan Road, Shanghai, 200240, China\\
  Key Laboratory for Particle Astrophysics and Cosmology (MOE),\\
  Shanghai Key Laboratory for Particle Physics and Cosmology (SKLPPC),\\
  Shanghai Jiao Tong University\\
  800 Dongchuan Road, Shanghai, 200240, China\\
  \texttt{zhiyuzhao@sjtu.edu.cn} \\
  \And
  Baohua Qi \\
  Institute of High Energy Physics, Chinese Academy of Sciences\\
  19B Yuquan Road, Beijing, 100049, China \\
  University of Chinese Academy of Sciences\\
  19A Yuquan Road, Beijing, 100049, China \\
  \texttt{qibh@ihep.ac.cn} \\
  \And
  Shu Li \\
  Tsung-Dao Lee Institute, Shanghai Jiao Tong University\\
  1 Lisuo Road, Shanghai, 201210, China \\
  Institute of Nuclear and Particle Physics, School of Physics and Astronomy\\
  800 Dongchuan Road, Shanghai, 200240, China\\
  Key Laboratory for Particle Astrophysics and Cosmology (MOE),\\
  Shanghai Key Laboratory for Particle Physics and Cosmology (SKLPPC),\\
  Shanghai Jiao Tong University\\
  800 Dongchuan Road, Shanghai, 200240, China\\
  \texttt{shuli@sjtu.edu.cn} \\
  \And
  Yong Liu \\
  Institute of High Energy Physics, Chinese Academy of Sciences\\
  19B Yuquan Road, Beijing, 100049, China \\
  University of Chinese Academy of Sciences\\
  19A Yuquan Road, Beijing, 100049, China \\
  \texttt{liuyong@ihep.ac.cn} \\
}

\renewcommand{\shorttitle}{\textit{arXiv} Template}

\hypersetup{
pdftitle={Dynamic Range of SiPMs with High Pixel Densities},
pdfauthor={Zhiyu Zhao},
pdfkeywords={SiPM, Dynamic range, BGO, Scintillator detector, Calorimeter, Higgs factory},
}

\begin{document}
\maketitle

\begin{abstract}
This study investigates the characteristics of Silicon Photomultipliers (SiPMs) with different pixel densities, focusing on their response across a wide dynamic range. Using an experimental setup that combines laser source and photomultiplier tubes (PMTs) for accurate light intensity calibration, we evaluated SiPMs with pixel counts up to 244,719 and pixel sizes down to 6 micrometers. To complement the experimental findings, a "Toy Monte Carlo" was developed to replicate the SiPMs' reponses under different lighting conditions, incorporating essential parameters such as pixel density and photon detection efficiency. The simulations aligned well with the experimental results for laser light, demonstrating similar nonlinearity trends. For BGO scintillation light, the simulations, which included multi-firing effect of pixels, showed significantly higher photon counts compared to the laser simulations. Furthermore, the simulated response derived in this research offer a method to correct for SiPM saturation effect, enabling accurate measurements in high-energy events even with SiPMs having a limited number of pixels.
\end{abstract}


\section{Introduction}
\label{sec:intro}
Silicon Photomultipliers (SiPMs) are at the forefront of photodetection technology, characterized by their high sensitivity, precision in light detection, counting, and measurement. These devices, composed of an array of avalanche photodiodes (APDs) operating in Geiger mode above the breakdown voltage, allow a single photon to initiate a self-sustaining avalanche. Noted for their high gain, superior timing resolution, low voltage requirements, and insensitivity to magnetic fields, SiPMs are extensively utilized in diverse fields, including medical imaging (notably in PET scans), LIDAR systems, astrophysical research, and high-energy physics. Their compact and robust design offers significant advantages over traditional photomultiplier tubes (PMTs), enabling a broader range of applications.

In high-energy collider experiments, calorimeters are essential detectors, playing a crucial role in precisely measuring particle energy. As the exploration of the structure of matter continues, there is an increasing demand for higher energy colliders and more precise detectors. Scintillation crystal calorimeters, which offer extremely high energy resolution due to their high light output, small statistical fluctuations, and high signal-to-noise ratios, have become more prevalent in large-scale colliders. Examples include the BGO electromagnetic calorimeter (ECAL)~\cite{SUMNER1988252} in the L3 experiment~\cite{ADEVA199035} at CERN's Large Electron-Positron Collider (LEP)~\cite{Myers:226776} and the PWO ECAL~\cite{CERN-LHCC-97-033} in the CMS experiment~\cite{CERN-LHCC-97-010} at the Large Hadron Collider (LHC)~\cite{Brüning:782076}. These calorimeters employ silicon photodiodes (SiPDs) and avalanche photodiodes (APDs), respectively, both of which have a large dynamic range to accommodate high light output. However, advancements in technology have led to the development of SiPMs with larger dynamic ranges, coupled with high gain and excellent time resolution, showing great potential for future collider calorimeter R\&D efforts\cite{CERN-DRDC-2024-004}.

The future Circular Electron-Positron Collider (CEPC) is a large international project aimed at precisely measuring the Higgs, W, and Z bosons, top quarks, and exploring Beyond Standard Model (BSM) physics~\cite{CEPCTDR-Accelerator, CEPCCDR-2}. A highly granular crystal ECAL has been proposed to address major challenges in jet reconstruction and achieve optimal electromagnetic energy resolution of around $2–3\%/\sqrt{E}$ with a homogeneous structure for CEPC~\cite{Liu_2020, instruments6030040}. In the conceptual design of this ECAL, long crystal bars, with BGO as an optional material, are arranged orthogonally between layers to achieve high granularity while minimizing the number of readout channels, with SiPMs placed at both ends of the crystal bars as optional photon sensors. At center-of-mass energies ranging from 240 GeV to 360 GeV, the energy deposition in a single crystal bar can reach up to 30 GeV, potentially resulting in the detection of approximately 350,000 photoelectrons per channel. This places significant demands on the dynamic range of SiPMs.

The dynamic range of SiPMs is inherently linked to the number of pixels they contain. Saturation occurs when the number of detected photons exceeds the available pixels, becoming more pronounced under high photon influx. Consequently, SiPMs with a higher pixel count exhibit delayed saturation effects, enhancing their utility in capturing extensive photon ranges. To fully harness this capacity, it is essential to understand the saturation threshold by mapping SiPM output across varying light intensities. Achieving this requires calibration against a scaling sensor capable of maintaining linear performance across the entire input spectrum, a challenging task given the constraints of available sensors.

The key issue is that the scaling sensor should maintain a linear output under strong light input while having sufficient gain to ensure precise calibration. Some experiments use SiPDs with very large dynamic ranges but almost no amplification~\cite{Tsuji:2020zsd}, or photomultiplier tubes with large amplification but small dynamic ranges~\cite{kotera2016}, as scaling sensors. However, the SiPMs tested in these experiments have relatively low pixel counts (less than 10 thousand).

We designed an experiment using a PMT to calibrate the number of incident photons on the SiPM. Typically, the PMT is more likely to saturate than the SiPM under the same light input. However, we can expand its linear region by increasing its bias voltage. By combining linear regions at different biases, we can achieve a linear region for the PMT with the highest possible gain. The dynamic range of three SiPMs with pixel sizes of 25 microns, 10 microns, and 6 microns was tested under laser illumination, all demonstrating a large dynamic range. Additionally, a model was developed to simulate the SiPMs' responses to both laser light and BGO crystal scintillation light.

\section{Experiment}
\label{sec:Experiment}

\subsection{Setup}
\begin{figure}[ht]
    \centering  
    \subfigure[]{
    \includegraphics[width=0.45\textwidth]{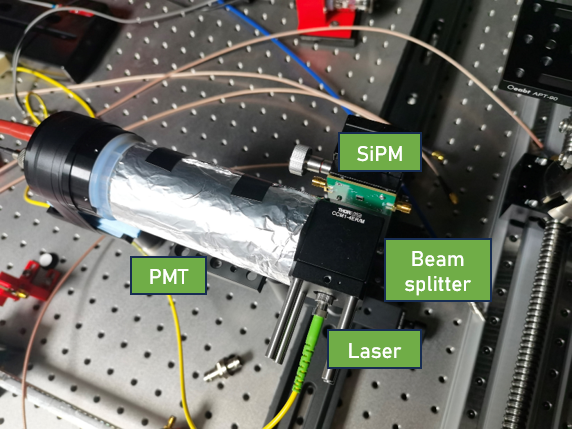}}
    \subfigure[]{
    \includegraphics[width=0.45\textwidth]{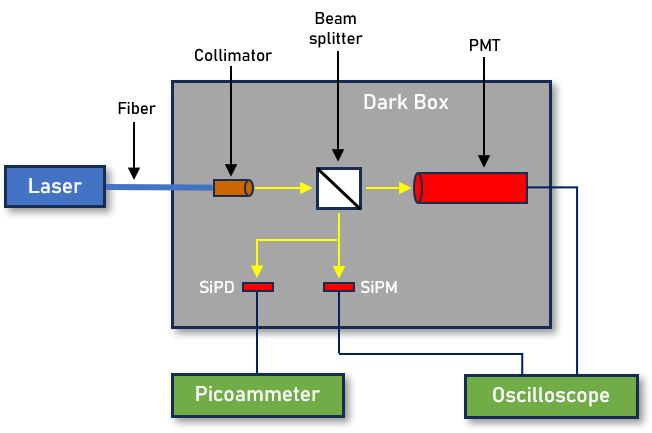}}
    \caption{\label{fig:Setup}~Diagrams of the experiment setup.}
\end{figure}

In Figure~\ref{fig:Setup}, a picosecond pulsed diode laser (NKT PIL040-FC~\cite{NKTPhotonics}, 405 nm, pulse width within 45 ps) serves as the light source. The emitted light is split into two beams using an optical beam splitter (or alternatively, an integrating sphere), directing one beam to the SiPM under test and the other to the PMT. We evaluated three different SiPM models: two from HAMAMATSU, one with a 25 $\mu$m pixel pitch and 57,600 pixels\cite{S13360-6025PE}, and another with a 10 $\mu$m pixel pitch and 89,984 pixels\cite{S14160-3010PS}, as well as one from NDL with a 6 $\mu$m pixel pitch and 244,719 pixels\cite{EQR06}. As listed in Table~\ref{tab:SiPM}, all featuring a high pixel count.

\begin{table*}[ht]
\centering
\caption{\label{tab:SiPM}~Properties of the measured SiPMs.}
    \begin{tabular}{ccccc}
        \toprule
        \makecell[c]{SiPM Model} &\makecell[c]{Pixel Pitch ($\mu$m)}  &\makecell[c]{Active Area (mm$^2$)} &\makecell[c]{Nominal pixel counts} &\makecell[c]{PDE (\%) \\ $\lambda=\lambda_p$}\\ 
        \midrule
        HAMAMATSU S13360-6025PE & 25 & 6.0$\times$6.0 & 57600 & 25\% \\
        HAMAMATSU S14160-3010PS & 10 & 3.0$\times$3.0 & 89984 & 18\% \\
        NDL EQR06 11-3030D-S & 6 & 3.0$\times$3.0 & 244719 & 30\% \\
        \bottomrule
    \end{tabular}
\end{table*}

As the intensity of the laser increases, the bias voltage of the PMT is correspondingly decreased. Since only a small fraction of the incident light reaches the PMT and the PMT operates under low bias voltages, its output remains linear across the entire input range, even when the SiPM reaches saturation. Consequently, the PMT's signal can be utilized to deduce the number of photons incident on the SiPM.

\subsection{SiPM calibration}
The gain of a SiPM is the amplification factor that increases the initial charge generated by a single photon striking the SiPM. This amplification is quantified by the number of charge carriers (electrons) produced per detected photon. Various factors, including temperature and bias voltage, can influence SiPM gain. Given the inherent variability among devices—even those of the same model—it is impractical to achieve absolute uniformity in gain across all units. As a result, individual calibration of each SiPM's gain under controlled conditions is necessary. For our experiments, temperature was regulated using the laboratory's air conditioning system, set to 25 degrees Celsius, while the bias voltage for each SiPM was kept constant, in accordance with the values specified in the manual.

\begin{figure}[ht]
    \centering  
    \subfigure[]{
    \includegraphics[width=0.45\textwidth]{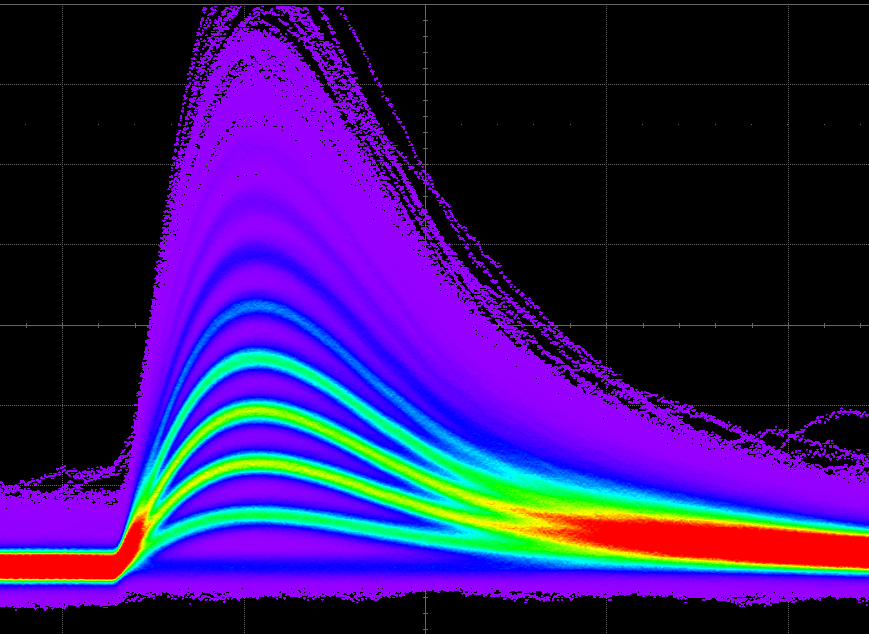}}
    \subfigure[]{
    \includegraphics[width=0.45\textwidth]{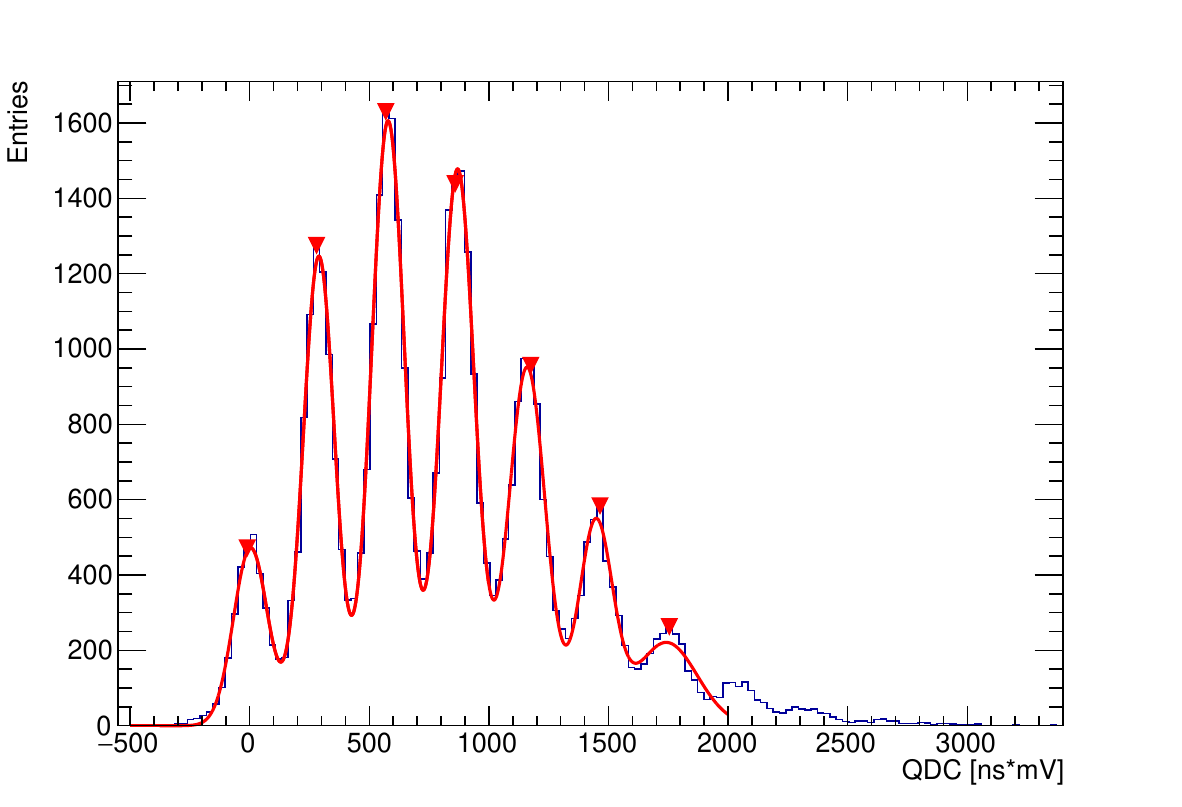}}
    \caption{\label{fig:SiPMCalibration}~(a) Waveforms output by SiPMs, where peaks of different amplitudes represent different numbers of p.e..  (b)QDC spectra of SiPMs, with a global Gaussian fitting applied to get peaks' positions.}
\end{figure}

Calibration was conducted under conditions of very weak incident light intensity, ensuring that each pulse emitted only a few photons, as depicted in Figure~\ref{fig:SiPMCalibration}(a). By integrating the waveform of the signal, we obtained the distribution of Charge-to-Digital-Converter (QDC) values (Figure~\ref{fig:SiPMCalibration}(b)), which correlates directly with the measured charge and is expressed in units of $ns\times mV$. The distinct peaks in Figure~\ref{fig:SiPMCalibration}(b) correspond to the charges associated with different photoelectron (p.e.) numbers. Linear regression applied to these peak positions allowed  us to calculate the QDC-to-p.e. ratio, indicating the QDC output per p.e.. To enhance the ability to distinguish signals from low photon counts, a pre-amplifier was employed. After calibrating for the amplification provided by the pre-amplifier, we determined the gains for three types of SiPMs, as shown in Table~\ref{tab:SiPMGain}.

\begin{table}[ht]
\centering
\caption{\label{tab:SiPMGain}~Calibrated QDC-to-p.e. ratios of SiPMs.}
    \begin{tabular}{ccc}
        \toprule
        \makecell[c]{S13360-6025PE} &\makecell[c]{S14160-3010PS}  &\makecell[c]{EQR06 11-3030D-S}\\ 
        \midrule
        6.42 ns$\cdot$mV/p.e. & 1.69 ns$\cdot$mV/p.e. & 0.9 ns$\cdot$mV/p.e. \\
        \bottomrule
    \end{tabular}
\end{table}

\subsection{PMT operation modes}
The phenomenon of saturation in PMTs arises due to the dynodes within each PMT, which serve to amplify the initial p.e. generated by an incident photon. As light intensity increases, the dynodes reach their amplification limit, unable to supply additional electrons for signal enhancement. This limitation results in a plateau in the PMT's output signal, although further increases in incoming light intensity, thereby causing saturation and constraining the PMT's dynamic range. However, reducing the PMT's voltage decreases its gain, thus reducing its susceptibility to saturation under identical light conditions. To optimize gain while preserving the PMT's linear response, we have devised several operational modes based on varying bias voltages. At low input light intensities, the PMT is set to operate in high bias mode, transitioning to low bias mode as light intensity increases.

\begin{figure}[ht]
    \centering
    \includegraphics[width=0.45\linewidth]{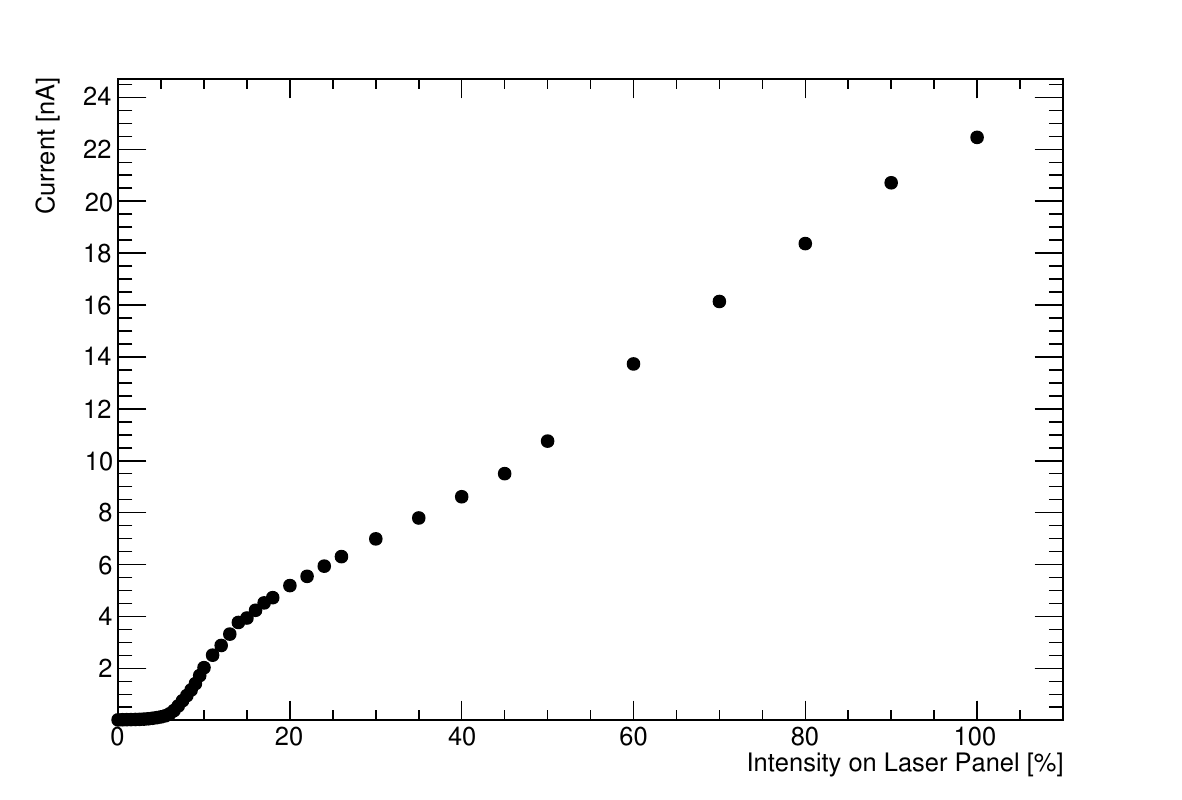}
    \caption{\label{fig:SiPDCurrent}~SiPD current variation with laser intensity.}
\end{figure}

\begin{figure}[ht]
    \centering 
    \subfigure[]{
    \includegraphics[width=0.45\textwidth]{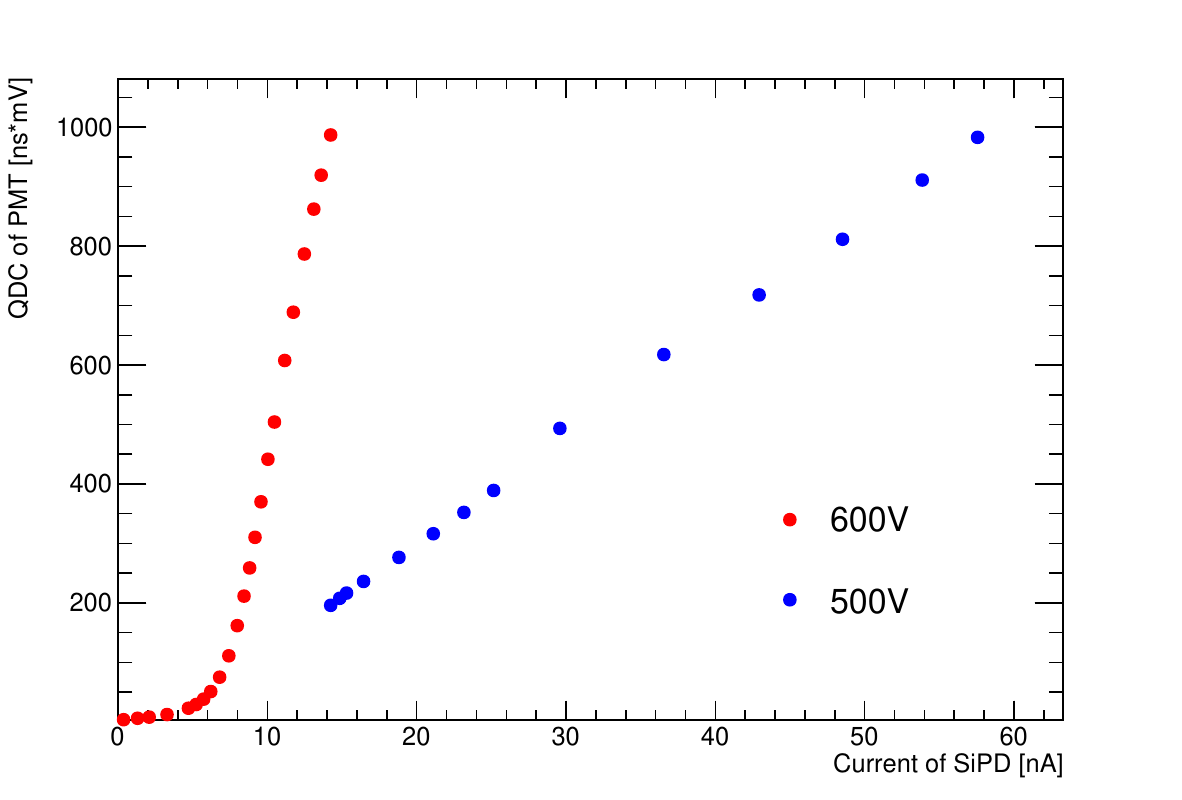}}
    \subfigure[]{
    \includegraphics[width=0.45\textwidth]{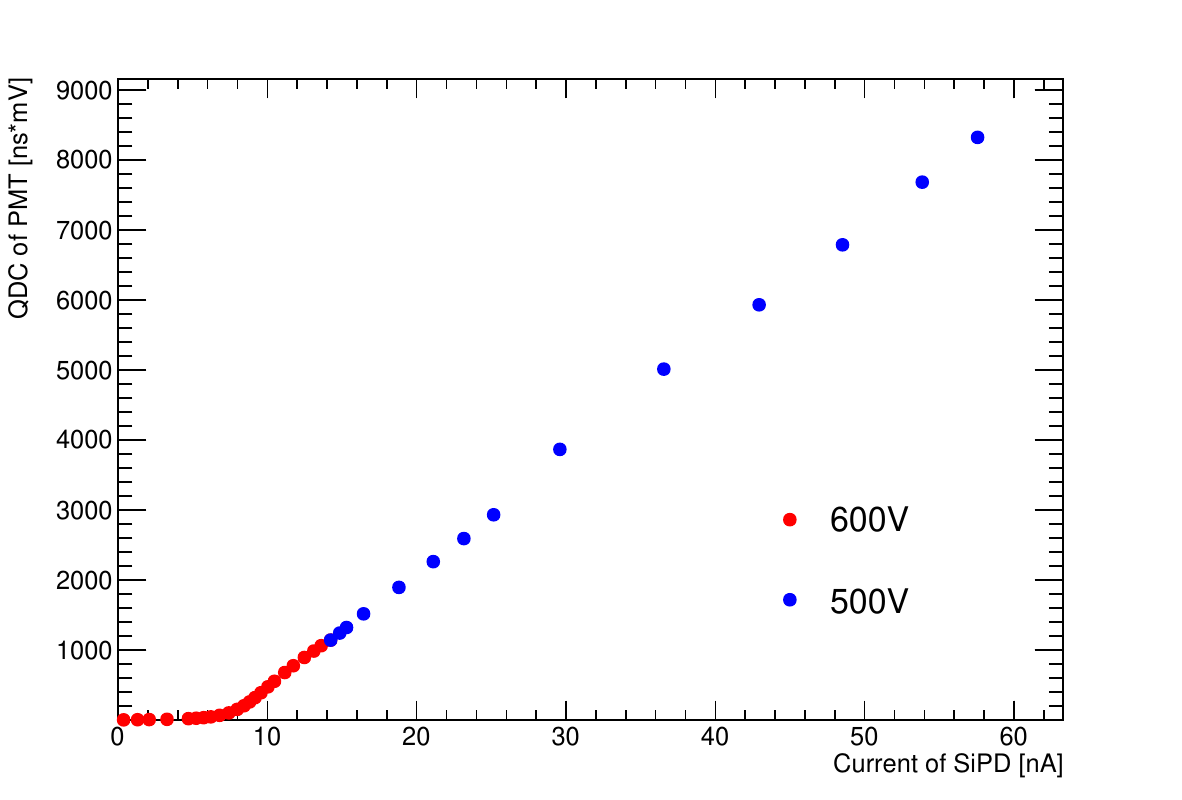}}
    \caption{\label{fig:PMTGainCalibtation}~PMT response at different bias voltages: (a) before gain calibration, (b) after gain calibration. The blue points represent PMT output at 500V bias voltage, while the red points represent PMT output at 600V bias voltage.}
\end{figure}

Prior to determining the optimal operational modes for the PMT, it is essential to calibrate the trend of light intensity changes received by the PMT. As shown in Figure~\ref{fig:Setup}(b), incident light from the laser is divided by a beam splitter, directing one beam towards the SiPD and the other towards the PMT. Given a constant splitter ratio, the SiPD's output current directly correlates with the photon count incident on the PMT's photocathode. Figure~\ref{fig:SiPDCurrent} illustrates the actual variance in laser intensity, which does not linearly match the set intensity. Initially, increasing the laser intensity elicits negligible response from the SiPD, which later transitions to a nearly linear response. Distinct bias voltages were selected for the PMT to accommodate different light intensities, with Figure~\ref{fig:PMTGainCalibtation} showing two linear response regions at 600V and 500V bias voltages.

\begin{figure}[ht]
    \centering 
    \subfigure[]{
    \includegraphics[width=0.45\textwidth]{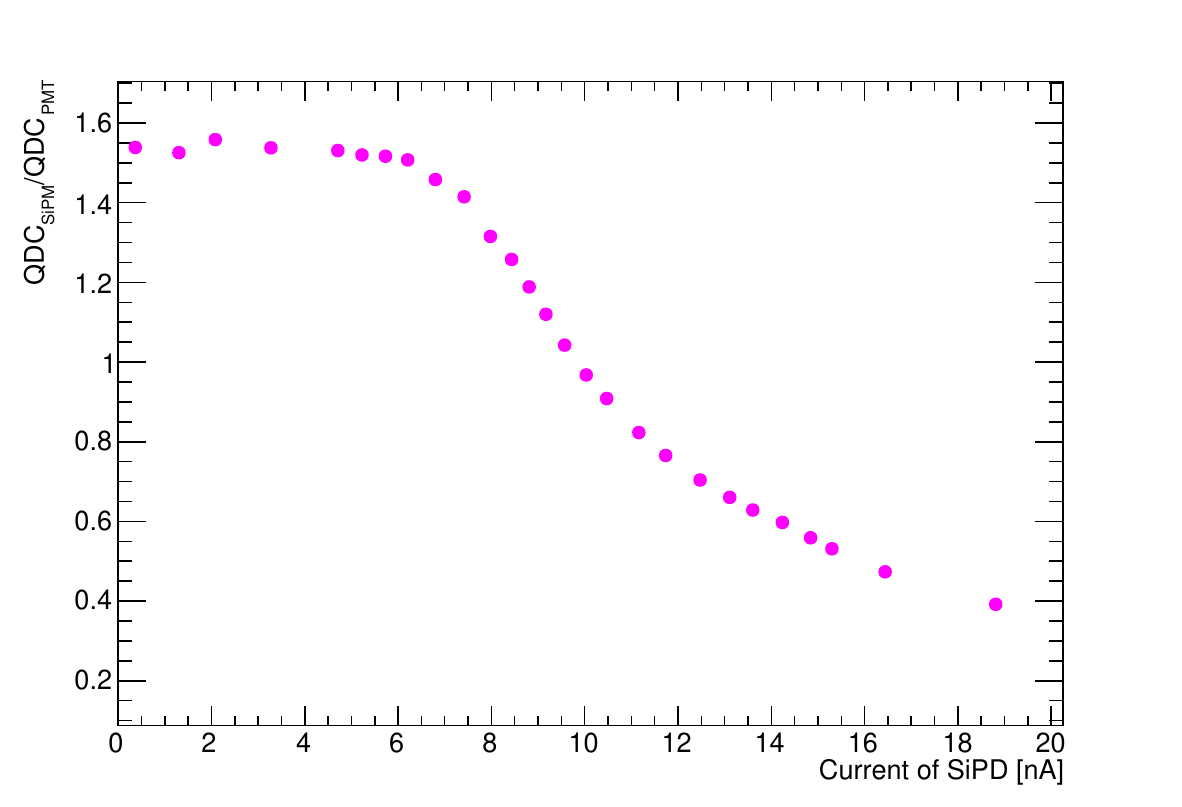}}
    \subfigure[]{
    \includegraphics[width=0.45\textwidth]{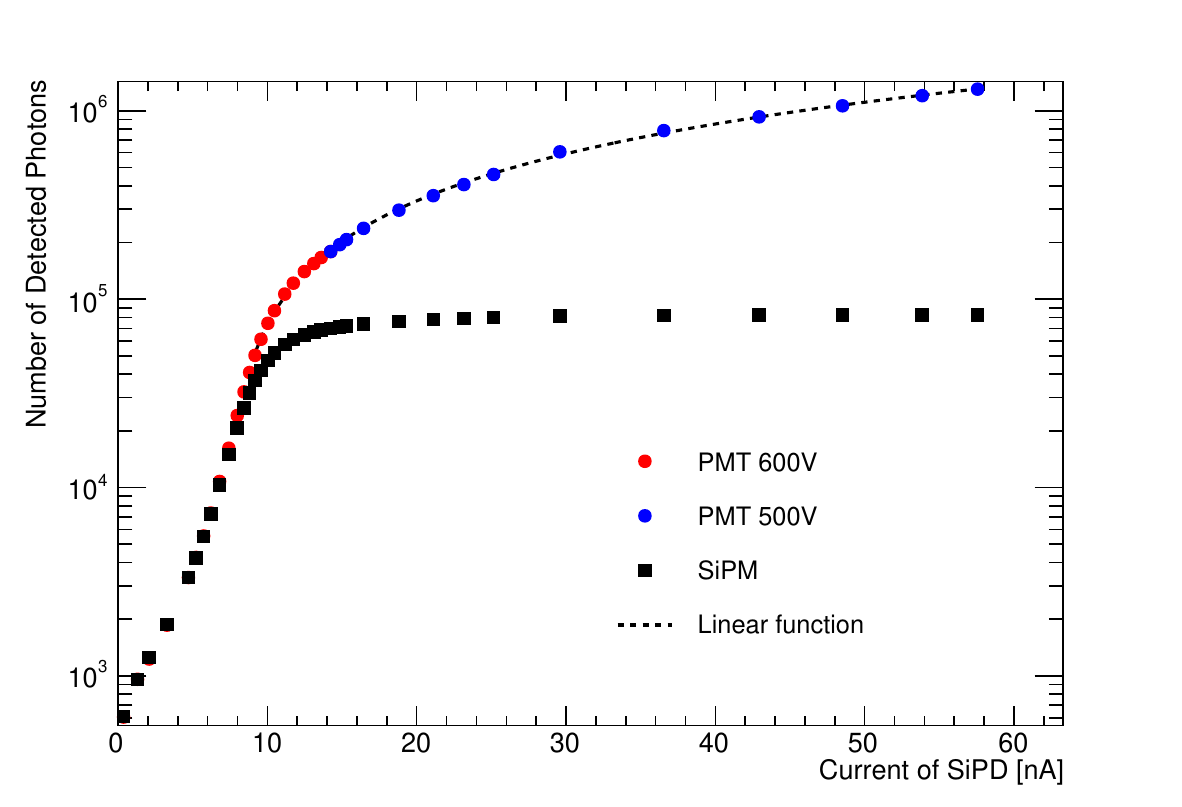}}
    \caption{\label{fig:PMTPE}~(a) Ratio of SiPM WDC to PMT QDC in weak light intensity region, initial a common linear response region both for SiPM and PMT. (b) Results after PMT calibration. The PMT can maintain linearity across the entire range, whereas the SiPM starts off linear but gradually deviates from linearity and eventually saturates.}
\end{figure}

The slopes of the data points in Figure~\ref{fig:PMTGainCalibtation}(a) indicate the PMT gains at respective bias voltages. The light intensity at their intersection is consistent. In Figure~\ref{fig:PMTGainCalibtation}(b), we adjusted the data for 500V to align its gain with that of 600V. Consequently, the PMT can maintain a linear response across nearly the entire range of laser intensity settings. Nonetheless, at a 600V bias voltage, the PMT's gain is too low to detect single photon signals, precluding direct calibration for single p.e. events. But in the weak light intensity region shown in Figure~\ref{fig:PMTPE}(a), the ratio of QDC measurements by SiPM and PMT remains nearly constant, indicating no saturation in either device. At these initial points, the photon count detected by the PMT can be equated to that detected by the SiPM. With increasing light intensity, the PMT maintains a linear response, allowing its output to facilitate the calibration of the expected photon count detected by the SiPM in the absence of saturation effects. In Figure~\ref{fig:PMTPE}(b), the laser, split by a beam splitter, simultaneously illuminates both the SiPM and PMT. In low-light intensity zones, the data from both devices align. As light intensity increases, the PMT's output continues to progress linearly, whereas the SiPM's output begins to saturate. Thus, after calibration, the PMT's output can represent the effective photon count that the SiPM would detect without saturation effects, equivalent to the product of the SiPM's Photon Detection Efficiency (PDE) and the incident photon count on the SiPM's surface.

\subsection{SiPM response}

\begin{figure*}[ht]
    \centering  
    \subfigure[]{
    \includegraphics[width=0.32\textwidth]{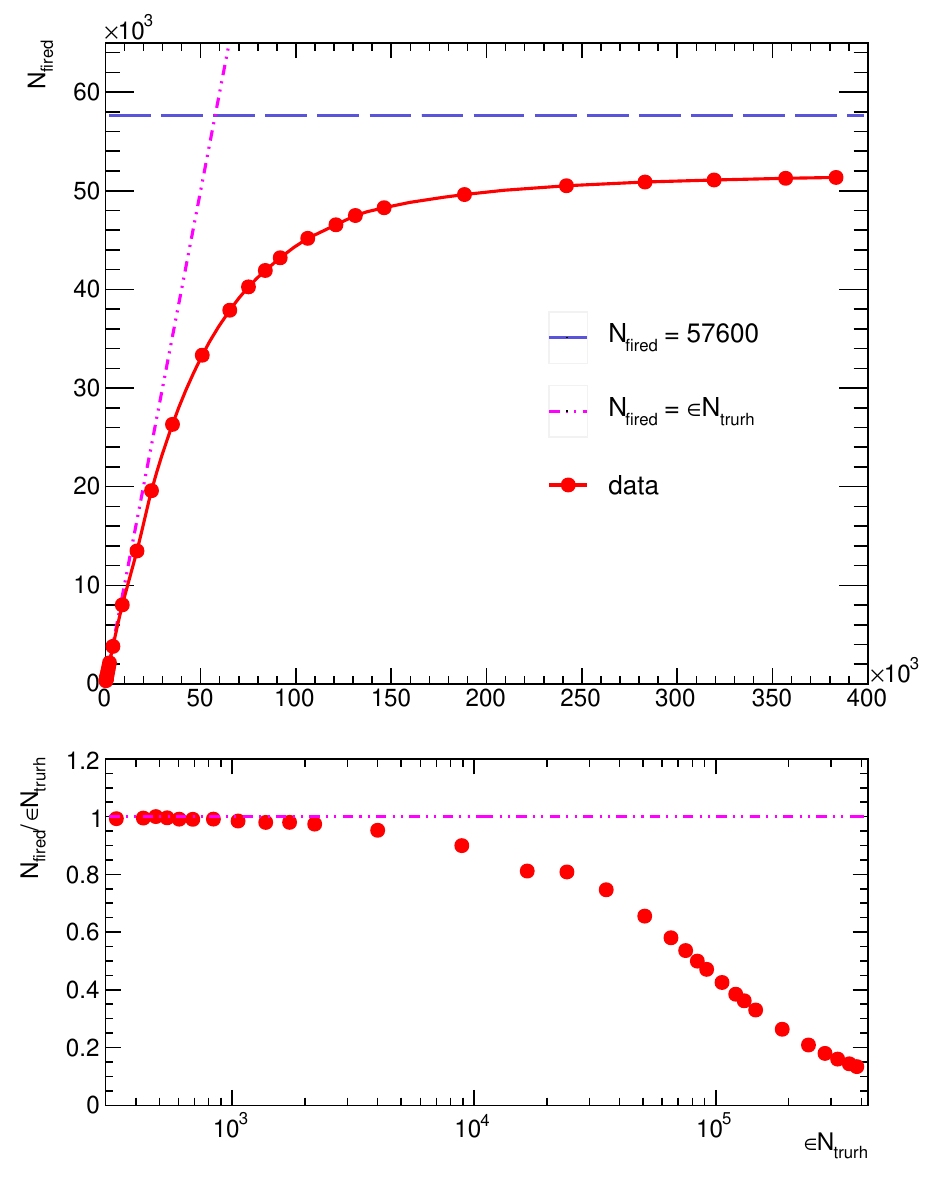}} 
    \subfigure[]{
    \includegraphics[width=0.32\textwidth]{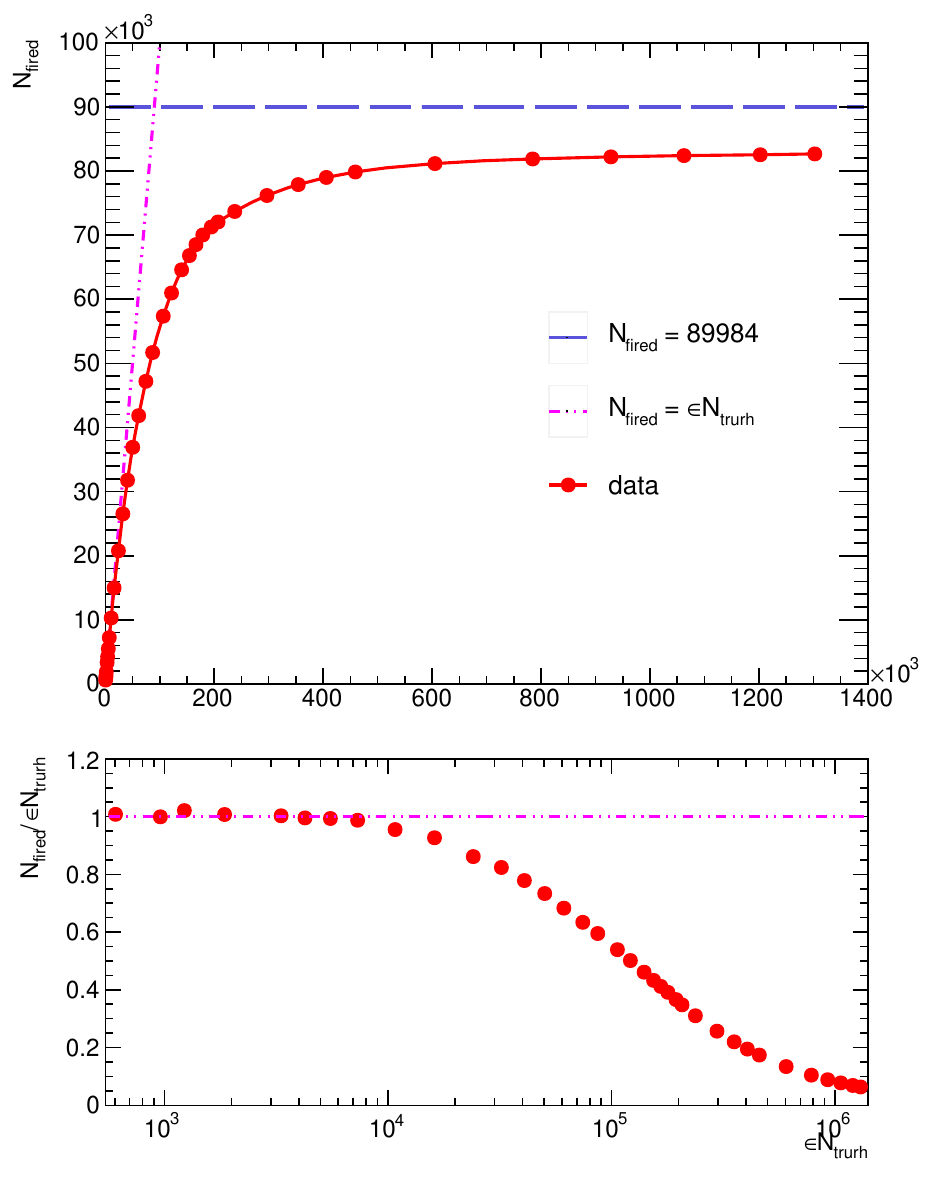}} 
    \subfigure[]{
    \includegraphics[width=0.32\textwidth]{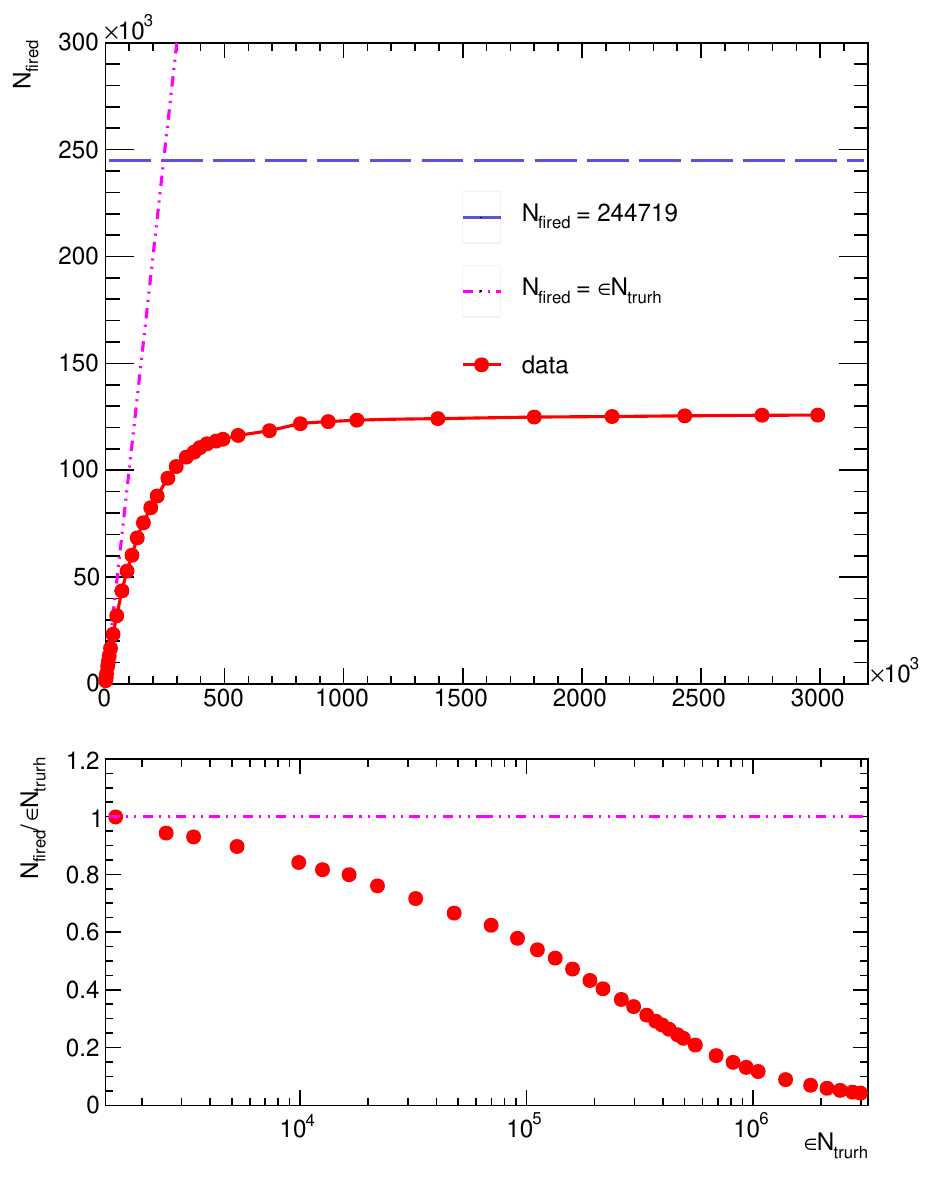}}
    \caption{\label{fig:SiPMCurve}~SiPM response as a function of the effective photon count: (a) HAMAMATSU S13360-6025PE (b) HAMAMATSU S14160-3010PS (c) NDL EQR06 11-3030D-S. In the upper figures, the red points represent the experimental test results, the blue dashed line represents the nominal pixel counts of the SiPM, and the purple dashed line is a slope of one. In the lower figures, the red points represent the ratio of actual photon counts to effective photon counts as measured by SiPM, with the purple dashed line equal to one.}
\end{figure*}

Figure~\ref{fig:SiPMCurve} shows the response of three types of SiPMs. The x-axis represents the PMT output, interpreted as the effective photon count. The red points indicate experimental test results, the blue dashed line represents the nominal pixel counts of the SiPM, and the purple dashed line is a slope of one, indicating that the number of photons detected by the SiPM equals the effective photon count. 

\begin{table}[ht]
\centering
\fontsize{7.5}{10}\selectfont
\caption{\label{tab:SiPMPixel}~Comparison between SiPMs' nominal pixel counts and their maximum photon counts under pico-second laser.}
    \begin{tabular}{ccc}
        \toprule
        \makecell[c]{SiPM} &\makecell[c]{Nominal pixel counts}  &\makecell[c]{Max. photon counts}\\ 
        \midrule
        S13360-6025PE & 57600  & 51347  \\
        S14160-3010PS & 89984  & 82664  \\
        EQR06 11-3030D-S & 244719 & 125775 \\
        \bottomrule
    \end{tabular}
\end{table}

The SiPM output initially exhibits a nearly linear increase, gradually deviates from linearity as the effective photon count increases, and eventually reaches a saturation region. For the two HAMAMATSU SiPMs with pixel counts of 57,600 and 89,984, nonlinearity effects start to appear at approximately 7\% to 10\% of the total pixel count. Their maximum output under laser illumination is close to, but slightly less than, the nominal pixel counts. Ideally, the saturation value should equal the pixel counts, since the experiments were conducted with a picosecond laser. However, due to manufacturing variations, the maximum photon counts detected by the SiPMs under these conditions may fluctuate. Table~\ref{tab:SiPMPixel} shows the maximum number of photons detected by the SiPMs under laser illumination and their actual pixel counts. 

For the SiPM with 244,719 pixels from NDL, the measured result does not meet expectations. The saturation value of this device is only about half of its nominal pixel count, while the nonlinearity effect starts in the very early region. Repeated experiments with the same device or other devices of the same type also showed similar performance. Further study is needed to understand these discrepancies.

\section{Simulation}
\label{sec:Simulation}

\subsection{SiPM response model}
To facilitate comparison with experimental results and to understand the saturation behavior of SiPMs under various conditions, we developed a SiPM response model. This model simulates the SiPM response, incorporating critical parameters such as pixel density, PDE, fill factor, avalanche triggering probability, crosstalk, and pixel recovery.

\begin{figure*}[ht]
    \centering
    \includegraphics[width=1.0\linewidth]{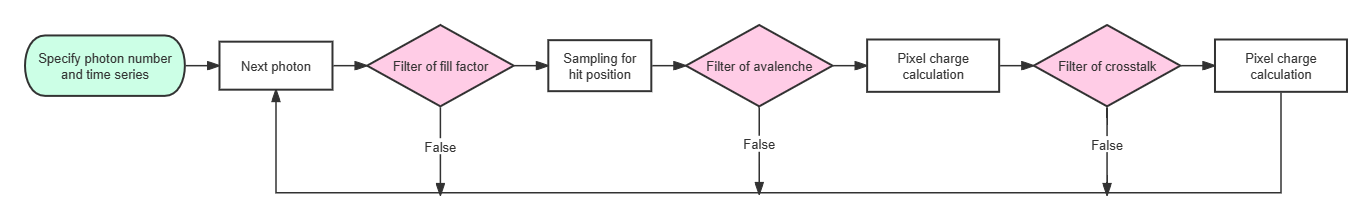}
    \caption{\label{fig:MCWorkflow}~Workflow of SiPM response model.}
\end{figure*}

The simulation workflow, depicted in Figure~\ref{fig:MCWorkflow}, begins with sampling the timestamp, wavelength, and incident position of each photon. These photons are then subjected to fill factor and avalanche probability filters based on their timestamps and wavelengths to determine if the currently impacted pixel will fire. A fired pixel generates a single-p.e. signal, and its response time is updated to the impact time. Another sampling is conducted to ascertain whether crosstalk occurs, which would result in the firing of an adjacent pixel.

In the simplest scenario, a single pixel will not re-fire and can only generate a single-p.e. signal. This requires all photons to arrive simultaneously or within a very short time interval at the SiPM. The laser tests described in Section~\ref{sec:Experiment} correspond to this situation, with the related simulation results presented in Section~\ref{subsec:MCLaser}. However, when the time intervals between incoming photons are longer, the response of a single pixel becomes more complex. A more refined simulation was introduced to describe its behavior, with the relevant results discussed in Section~\ref{subsec:MCBGO}.

\subsection{SiPM response to laser}
\label{subsec:MCLaser}
In this part, to compare with the experiment, the parameters were set as closely as possible to the experimental values. The PDE of the SiPM in the simulation was set to the value for 405 nm light, and it was assumed that a single pixel, once fired, would not re-fire, because the pulse width of the laser used in the experiment was less than 45 ps, much shorter than the time required for a SiPM pixel to re-fire.

\begin{figure*}[ht]
    \centering  
    \subfigure[]{
    \includegraphics[width=0.32\textwidth]{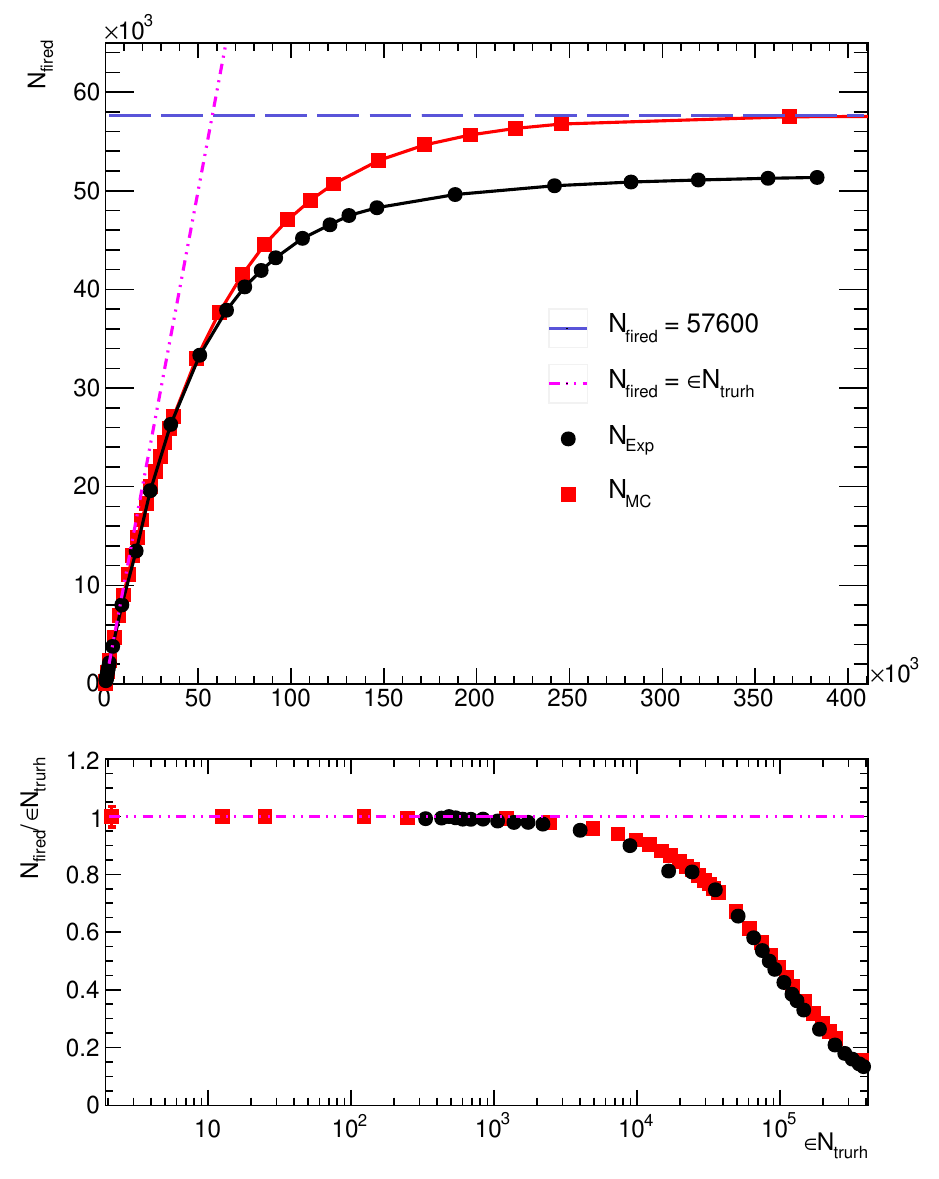}} 
    \subfigure[]{
    \includegraphics[width=0.32\textwidth]{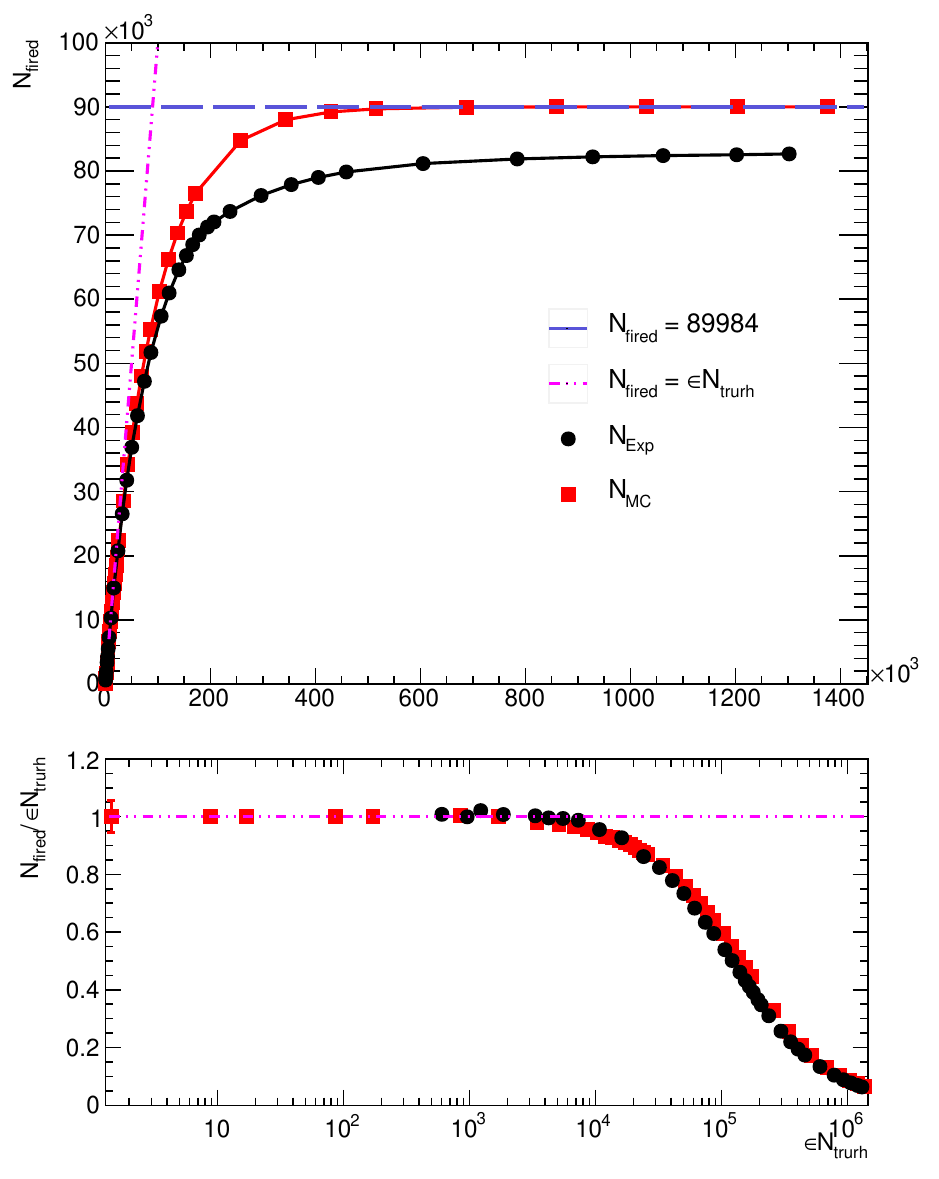}} 
    \subfigure[]{
    \includegraphics[width=0.32\textwidth]{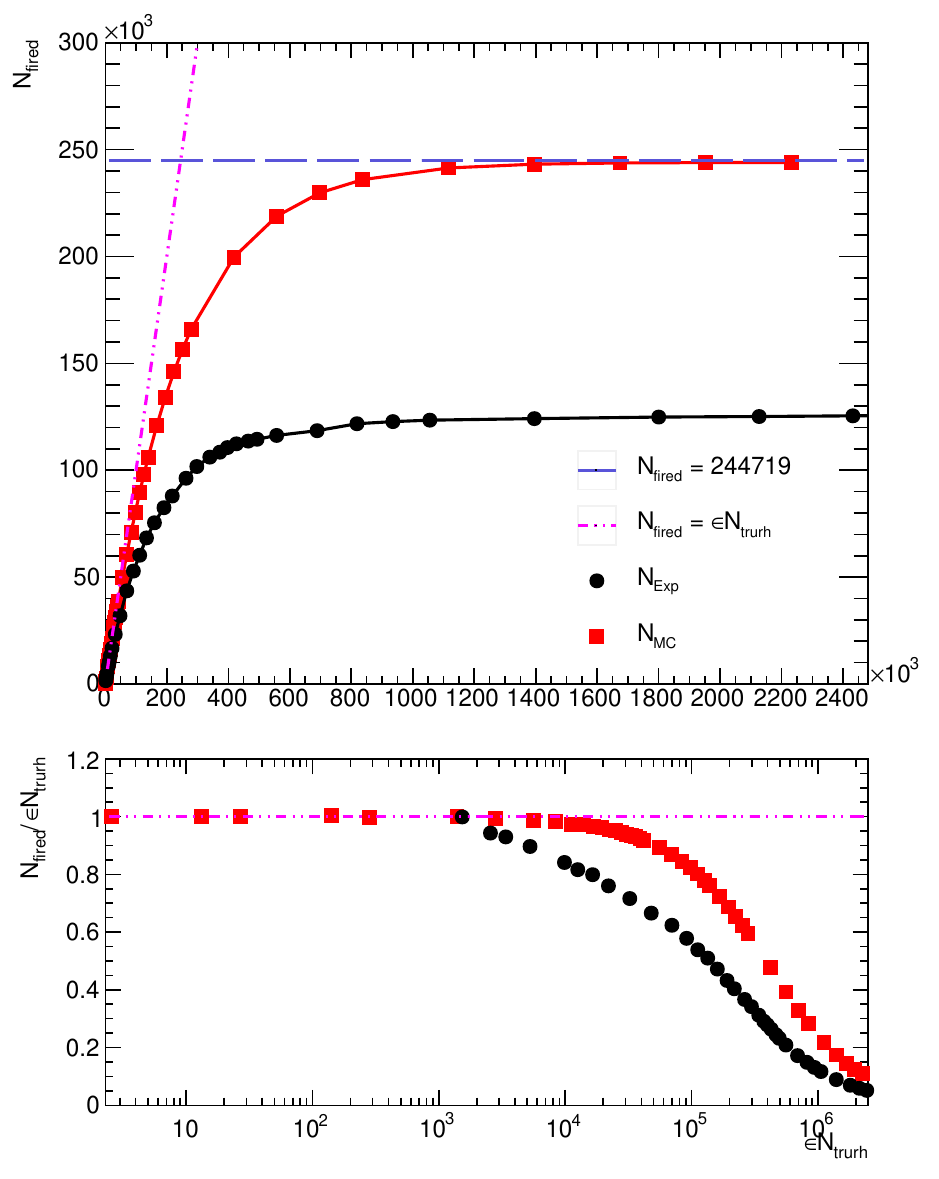}}
    \caption{\label{fig:SiPM_noRec}~Simulated response for three types of SiPMs and their comparison with the experimental measurements: (a) HAMAMATSU S13360-6025PE (b) HAMAMATSU S14160-3010PS (c) NDL EQR06 11-3030D-S. In the upper figures, the black points represent the experimental test results, the red points represent the simulated results, the blue dashed line represents the nominal pixel counts of the SiPM, and the purple dashed line is a slope of one. In the lower figures, the black and red points represent the ratio of actual photon counts to effective photon counts as measured by SiPM for experimental data and simulation, respectively, with the purple dashed line equal to one.}
\end{figure*}

\begin{table}[ht]
\centering
\caption{\label{tab:SiPMSatPoint_Laser}~Effective photon counts at the deviation point of 5\% nonlinearity using simulated results, measuring light from a picosecond laser.}
    \begin{tabular}{ccc}
        \toprule
        \makecell[c]{S13360-6025PE} &\makecell[c]{S14160-3010PS}  &\makecell[c]{EQR06 11-3030D-S} \\
        \midrule
        6049 & 9578 & 25381 \\
        \bottomrule
    \end{tabular}
\end{table}

Figure~\ref{fig:SiPM_noRec} shows the simulation results for three types of SiPMs and their comparison with the experimental measurements. As expected, the saturation values of these simulated results at very high light intensities are exactly equal to their respective pixel counts, since each pixel was assumed to be functioning normally in the simulation. Table~\ref{tab:SiPMSatPoint_Laser} lists the effective photon count corresponding to 5\% nonlinearity for the three SiPMs, with simulated results.

For the two Hamamatsu SiPMs with 57,600 and 89,984 pixels, although the measured saturation values are slightly lower than the simulated results, the trends of nonlinearity are very close to the experimental data, with deviations starting at approximately the same points. However, for the NDL EQR06 11-3030D-S with 244,719 pixels, there is still a significant difference between the simulation and the experimental results, indicating that the response of this device cannot be described by the general model presented in this paper.

\subsection{SiPM response to scintillator}
\label{subsec:MCBGO}
When measuring the light emitted by scintillators using SiPMs, it is crucial to account for the multiple firing effects of SiPM pixels. Given the decay time of scintillator emissions, which can range from tens to hundreds of nanoseconds for many scintillators, pixels in the SiPM that have been fired can be re-fired by incident photons after discharging. This re-fired can occur multiple times, depending on the duration of the decay time. A detailed model was built to describe the response of SiPMs to the scintillation light from BGO crystals, incorporating the multiple firing effects of SiPM pixels.

\subsubsection{Avalanche and recovery of SiPM pixels}
\label{subsubsec:Avalanche}
SiPMs consist of numerous pixels, each featuring a photodiode and a quenching resistor, which work together to detect photons through an avalanche multiplication process. This process, initiated by a single photon's interaction with the photodiode, results in a detectable current pulse. The recovery effect in SiPMs refers to the process by which a pixel within the SiPM becomes ready to detect another photon after it has previously detected one. The recovery time, the period it takes for a pixel to recharge and regain its initial state, significantly impacts the device's linearity and dynamic range, particularly in environments of intense photon input.

\begin{figure}[ht]
    \centering  
    \subfigure[]{
    \includegraphics[width=0.45\textwidth]{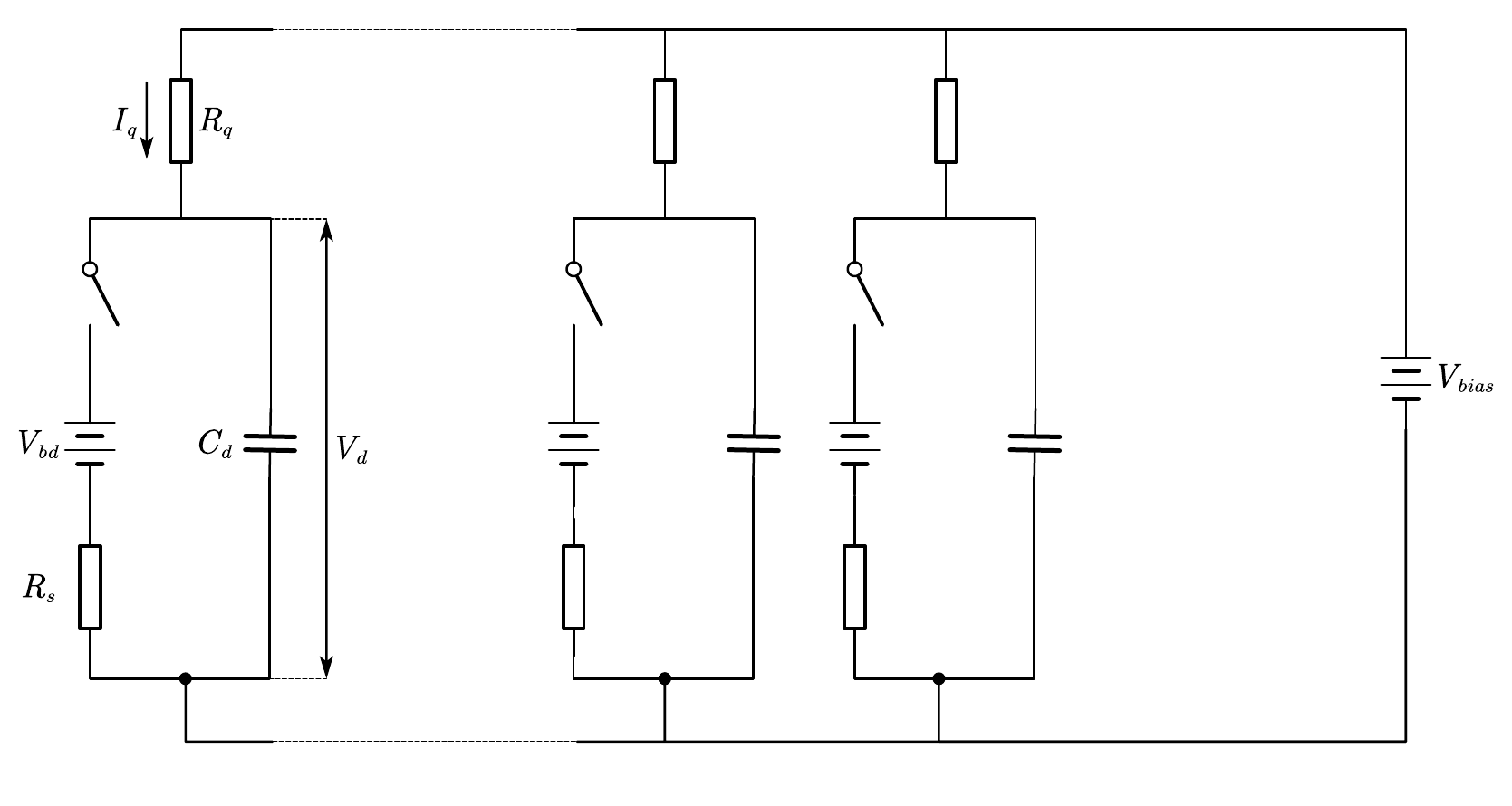}}
    \subfigure[]{
    \includegraphics[width=0.45\textwidth]{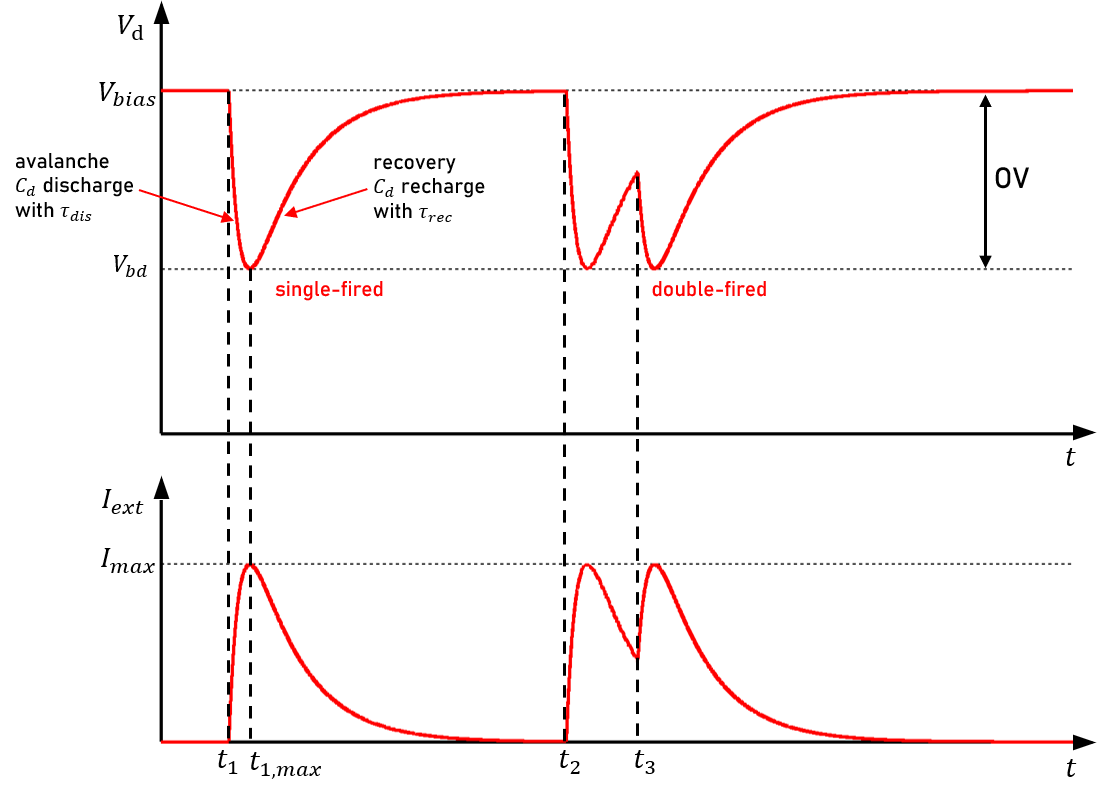}}
    \caption{\label{fig:Microcell}~(a) SiPM electronic model (b) Pixel operation in single-firing and double-firing events.}
\end{figure}

Significant research has been devoted to understanding the operating principles of SiPMs~\cite{NAGY201444, Piemonte:2019kll}. Figure~\ref{fig:Microcell} illustrates the electronic model of an SiPM and how a pixel operates under single-firing and double-firing events. An avalanche event leads to the discharging of the junction capacitance ($C_d$) marked by a time constant $\tau_{dis}$, followed by the recharging process, an exponential recovery of the voltage across the junction ($V_d$) to its bias level ($V_{bias}$), marked by a recovery time constant $\tau_{rec}$. This recovery process is pivotal in preparing the pixel for subsequent photon detections.

The recovery voltage $V_d$ for a complete single photon signal is given by:
\begin{equation}
V_d(t)=V_0-V_1\cdot(1-e^{-\frac{t}{\tau_{dis}}}) \cdot e^{-\frac{t}{\tau_{rec}}}
\end{equation}
where $V_0$ represents the initial voltage at the start of the avalanche, equating to $V_{bias}$ upon full signal recovery. $V_1$ is a parameter ensuring the function's minimum value aligns with $V_{bd}$. 

The waveform of the output signal by an SiPM pixel, correlating with the pixel's charge and discharge cycles of $C_d$, is encapsulated by:
\begin{equation}
I_{ext}(t)=I_0\cdot(1-e^{-\frac{t}{\tau_{dis}}}) \cdot e^{-\frac{t}{\tau_{rec}}}
\end{equation}
Signal amplitude is modulated by $I_0$, directly proportional to the overvoltage (OV, $V_{bias}$ - $V_{bd}$). Arrival of a subsequent photon during a pixel's incomplete recovery results in a reduced $V_d$, affecting $I_0$:
\begin{equation}
I_0(t)=I_{1}\cdot(1-e^{-\frac{t}{\tau_{rec}}})
\end{equation}
Adjusting $I_1$ can control the maximum amplitude that a single pixel can output.

The probability of avalanche triggering, $P_{trig}$, corresponds to the likelihood of a carrier within a pixel initiating the avalanche process. This probability scales with OV and depends on the pixel's state during recovery:
\begin{equation}
P_{trig}=P_{0}\cdot(1-e^{-\frac{t}{\tau_{rec}}})
\end{equation}
where $P_{0}$ denotes the trigger probability in a fully recharged pixel, subject to specific conditions such as OV, temperature, and wavelength.

\subsubsection{Photon tracing}
\label{subsubsec:PhotonTracing}
To accurately account for the multiple firing effect in SiPM applications, it is essential to determine the wavelength and timestamp of each photon's arrival at the SiPM surface. This information is obtained through a Geant4 optical simulation~\cite{Geant4}. Geant4 is a comprehensive software toolkit designed for simulating the passage of particles through matter, which can effectively model the energy absorption and subsequent light emission by scintillators, as well as the propagation of light through various materials.

\begin{figure}[ht]
    \centering
    \includegraphics[width=0.7\linewidth]{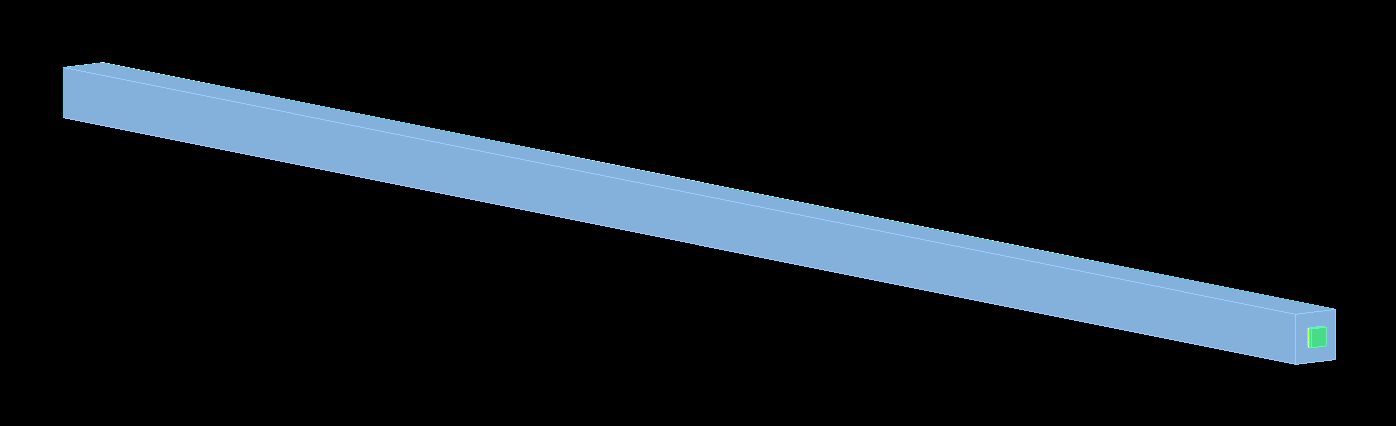}
    \caption{\label{fig:G4Optcal}~Geant4 simulation of a BGO crystal bar with dimensions of 40 $\times$ 40 $\times$ 1 cm$^3$, covered with ESR film, and SiPMs positioned at both ends.}
\end{figure}

In the simulation, a BGO crystal unit was constructed, comprising a BGO scintillating crystal enclosed in a reflective ESR film, with a SiPM positioned at both ends to detect scintillation photons (Figure~\ref{fig:G4Optcal}). A 20 $\mu$m air gap separates all materials. The properties of the BGO crystal and reflective film are detailed in Tables~\ref{tab:CrystalProperty} and~\ref{tab:FilmProperty}.

\begin{table}[ht]
\centering
\fontsize{7.5}{11}\selectfont
\caption{\label{tab:CrystalProperty}~Properties of BGO crystal.}
    \begin{tabular}{ccccc}
        \toprule
        \makecell[c]{Crystal} &\makecell[c]{Volume}  &\makecell[c]{Density} &\makecell[c]{Light Yield} &\makecell[c]{Decay time}\\
        \midrule
        BGO & 1$\times$1$\times$40 cm$^3$ & 7.13 g/cm$^3$ & 8200 pho/MeV &  60/300 ns \\
        \bottomrule
    \end{tabular}
\end{table}

\begin{table}[ht]
\centering
\caption{\label{tab:FilmProperty}~Properties of reflective film.}
    \begin{tabular}{cccc}
        \toprule
        \makecell[c]{Film} &\makecell[c]{Thickness($\mu$m)}  &\makecell[c]{Density(g/cm$^3$)} &\makecell[c]{Reflectivity\\$\lambda=\lambda_p$}\\
        \midrule
        ESR & 50 & 0.9 & 0.98 \\
        \bottomrule
    \end{tabular}
\end{table}

\begin{figure}[ht]
    \centering
    \subfigure[]{
    \includegraphics[width=0.45\textwidth]{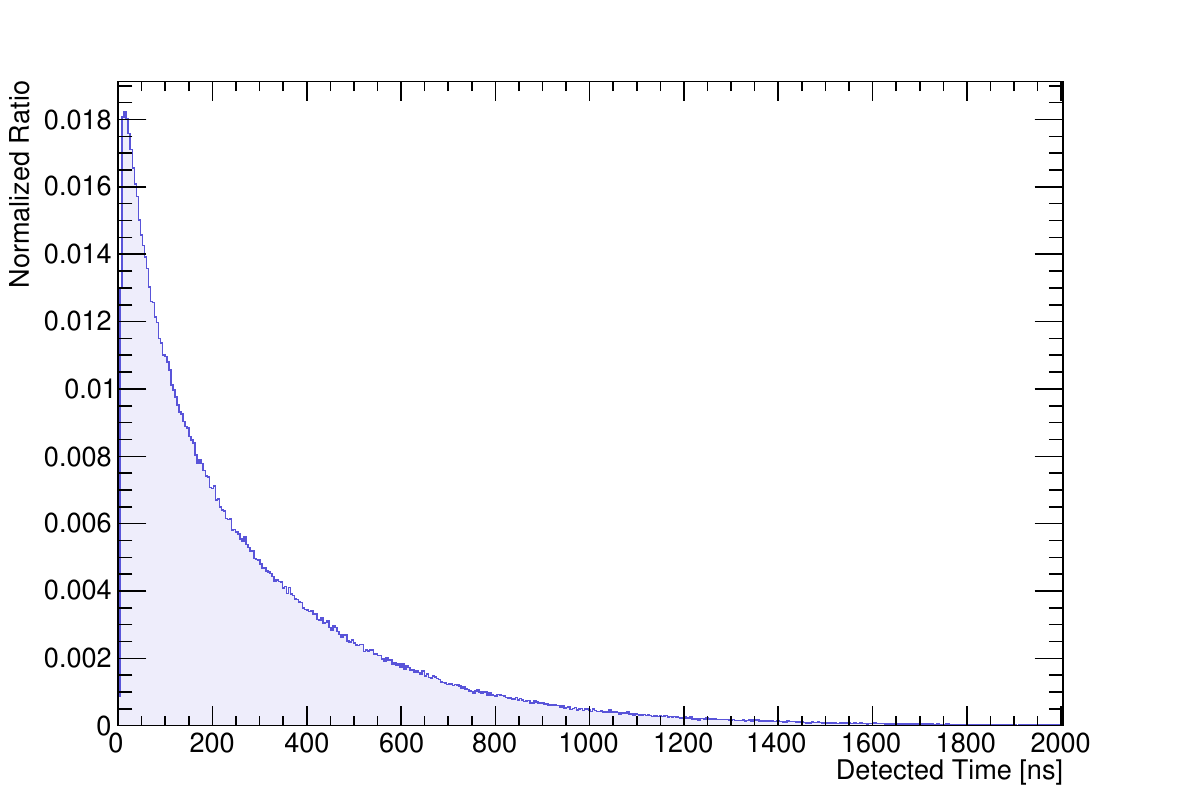}}
    \subfigure[]{
    \includegraphics[width=0.45\textwidth]{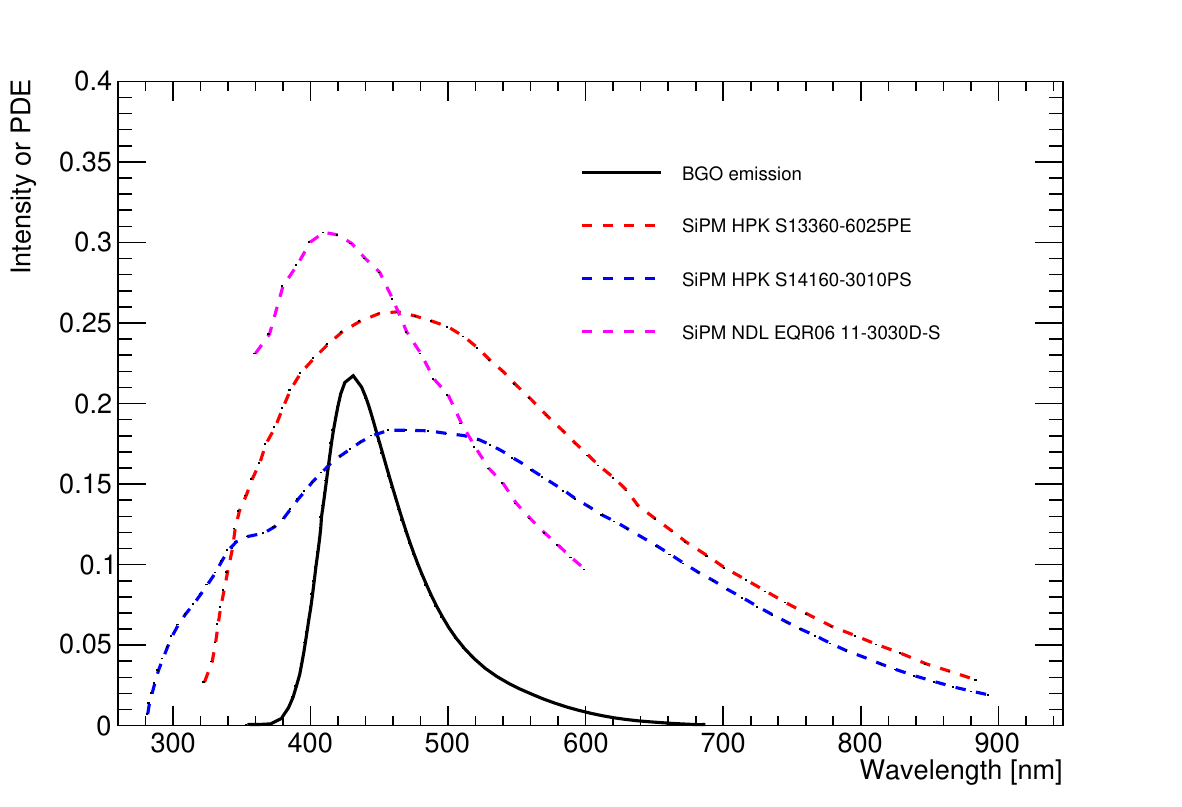}}
    \caption{\label{fig:TimeWavelength}~(a) Time distribution of BGO scintillation light detected by the SiPM. (b) BGO emission spectrum and SiPM PDE spectrum.}
\end{figure}

During the simulation, a high-energy particle impacts the BGO scintillator, depositing energy that results in light emission. The emitted photons undergo various optical interactions within the medium, including reflection, refraction, and absorption. Some of these photons may reach the SiPMs located at both ends of the crystal, either directly or after multiple reflections. The wavelengths and arrival times of these photons are recorded for sampling (Figure~\ref{fig:TimeWavelength}).

\subsubsection{Waveform parameters}
The output waveform of the SiPM can be regarded as the cumulative result of multiple single-photoelectron waveforms. Therefore, by understanding the single-photoelectron waveform, and combining the response function discussed in Section~\ref{subsubsec:Avalanche} and the detection times and wavelengths of the photons covered in Section~\ref{subsubsec:PhotonTracing}, the actual output waveform of the SiPM can be obtained through superposition.

\begin{figure}[ht]
    \centering  
    \subfigure[]{
    \includegraphics[width=0.45\textwidth]{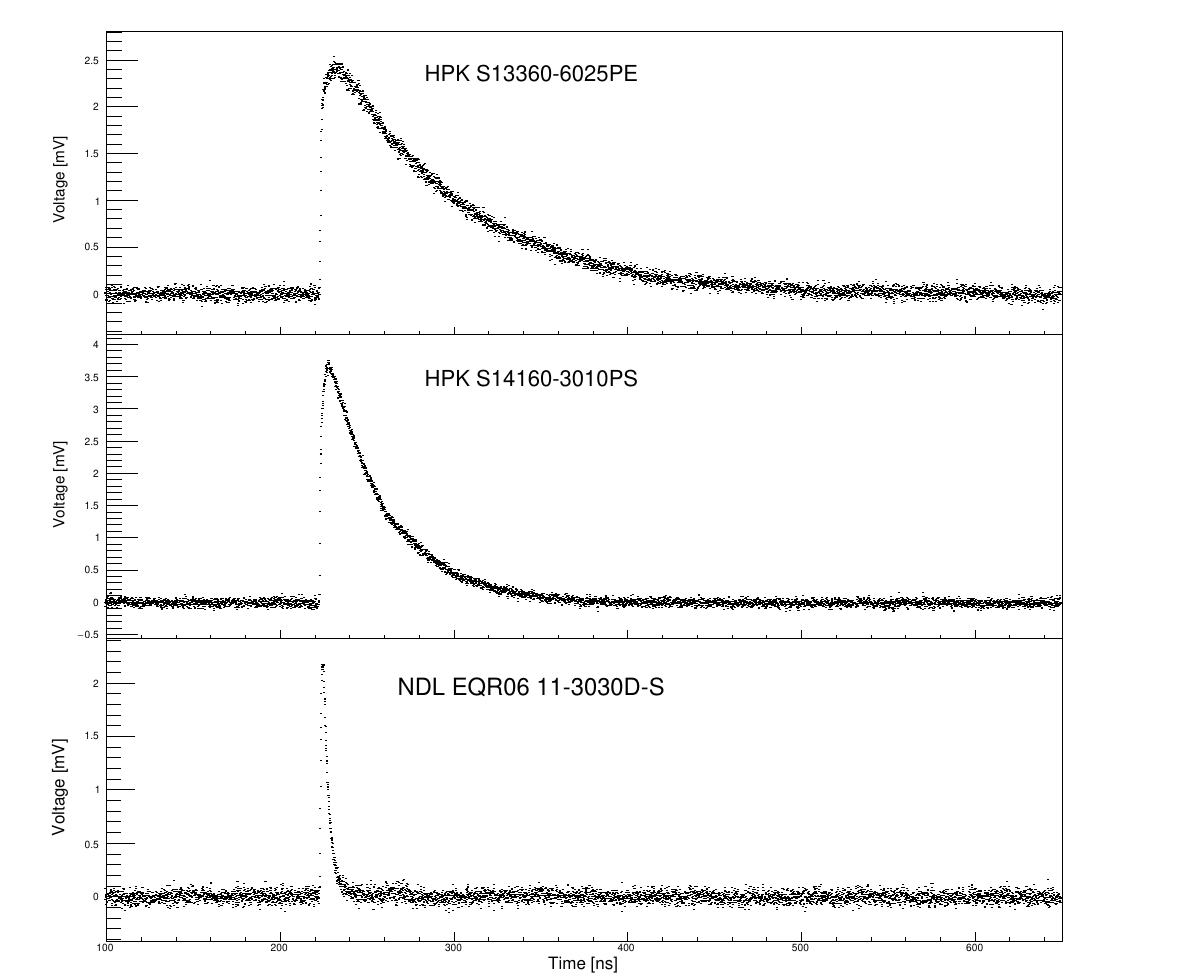}}
    \subfigure[]{
    \includegraphics[width=0.45\textwidth]{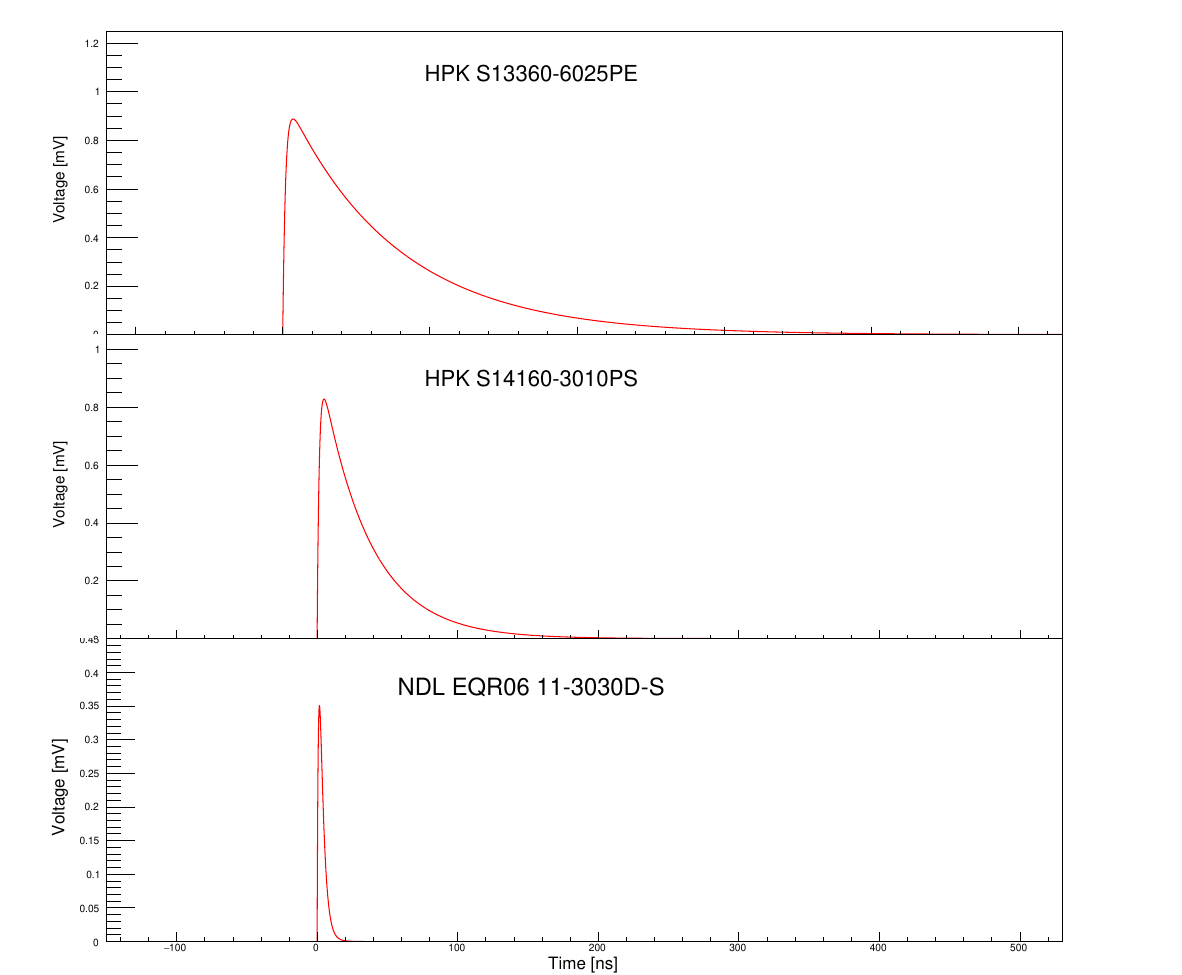}}
    \caption{\label{fig:WaveformSimu}~(a) Experimentally measured waveforms of the SiPM, obtained using a picosecond laser operating at low intensity. (b) Simulated SiPM waveforms, based on the derived waveform parameters.}
\end{figure}

The mathematical expression for a single-photoelectron waveform is given by:
\begin{equation}
A(t)=A_0\cdot(1-e^{-\frac{t}{\tau_{dis}}}) \cdot e^{-\frac{t}{\tau_{rec}}}
\label{eq:SiPMWaveform}
\end{equation}
where $A_0$ represents the amplitude parameter, influenced by the pixel's recovery state. Figure~\ref{fig:WaveformSimu}(a) presents the SiPM waveforms obtained using a laser at low intensity as the light source. To avoid bandwidth limitations, no signal amplifier was employed in the experiment. These waveforms may comprise hundreds of p.e.. However, given the picosecond-scale pulse width of the laser, the features of these waveforms should closely resemble those of single-p.e. waveforms, differing primarily in amplitude. The waveforms of the three SiPMs exhibit noticeable differences. Generally, as the pixel size of the SiPM decreases, the waveform broadens. By fitting the waveform in Figure~\ref{fig:WaveformSimu}(a) using Equation~\ref{eq:SiPMWaveform}, the characteristic time constants of their leading and falling edges can be determined, as shown in Table~\ref{tab:WaveformPar}. The simulated single-p.e. waveforms are shown in Figure~\ref{fig:WaveformSimu}(b).

\begin{table}[ht]
\centering
\fontsize{7.5}{11}\selectfont
\caption{\label{tab:WaveformPar}~Waveform parameters of SiPMs.}
    \begin{tabular}{cccc}
        \toprule
        \makecell[c]{SiPM} &\makecell[c]{S13360-6025PE}  &\makecell[c]{S14160-3010PS} &\makecell[c]{EQR06 11-3030D-S}\\
        \midrule
        $\tau$$_{dis}$ (ns) & 1.91 & 1.58 & 1.66\\
        $\tau$$_{rec}$ (ns) & 74.64 & 34.22 & 2.82\\
        \bottomrule
    \end{tabular}
\end{table}

\subsubsection{Results and discussion}
\begin{figure*}[ht]
\centering  
\subfigure[]{
\includegraphics[width=0.32\textwidth]{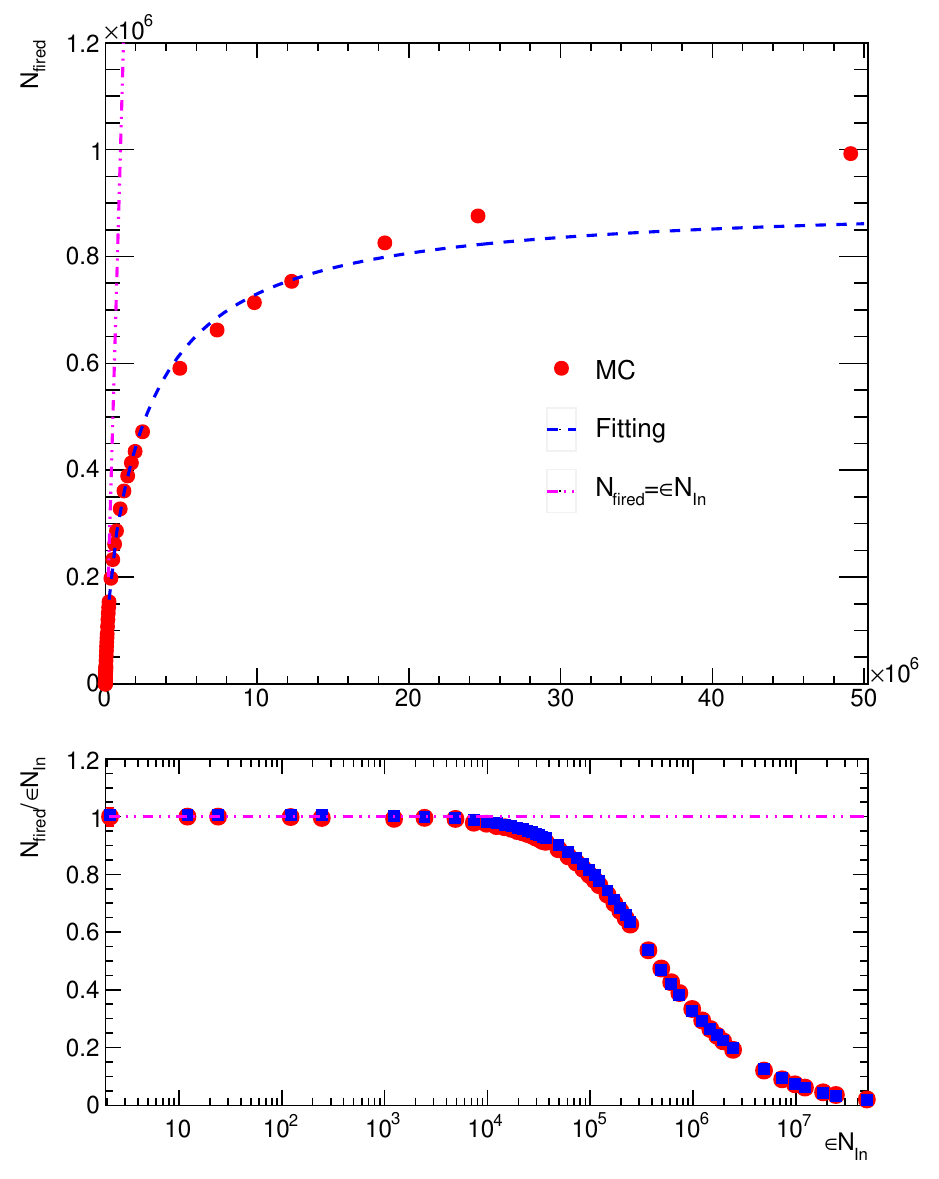}} 
\subfigure[]{
\includegraphics[width=0.32\textwidth]{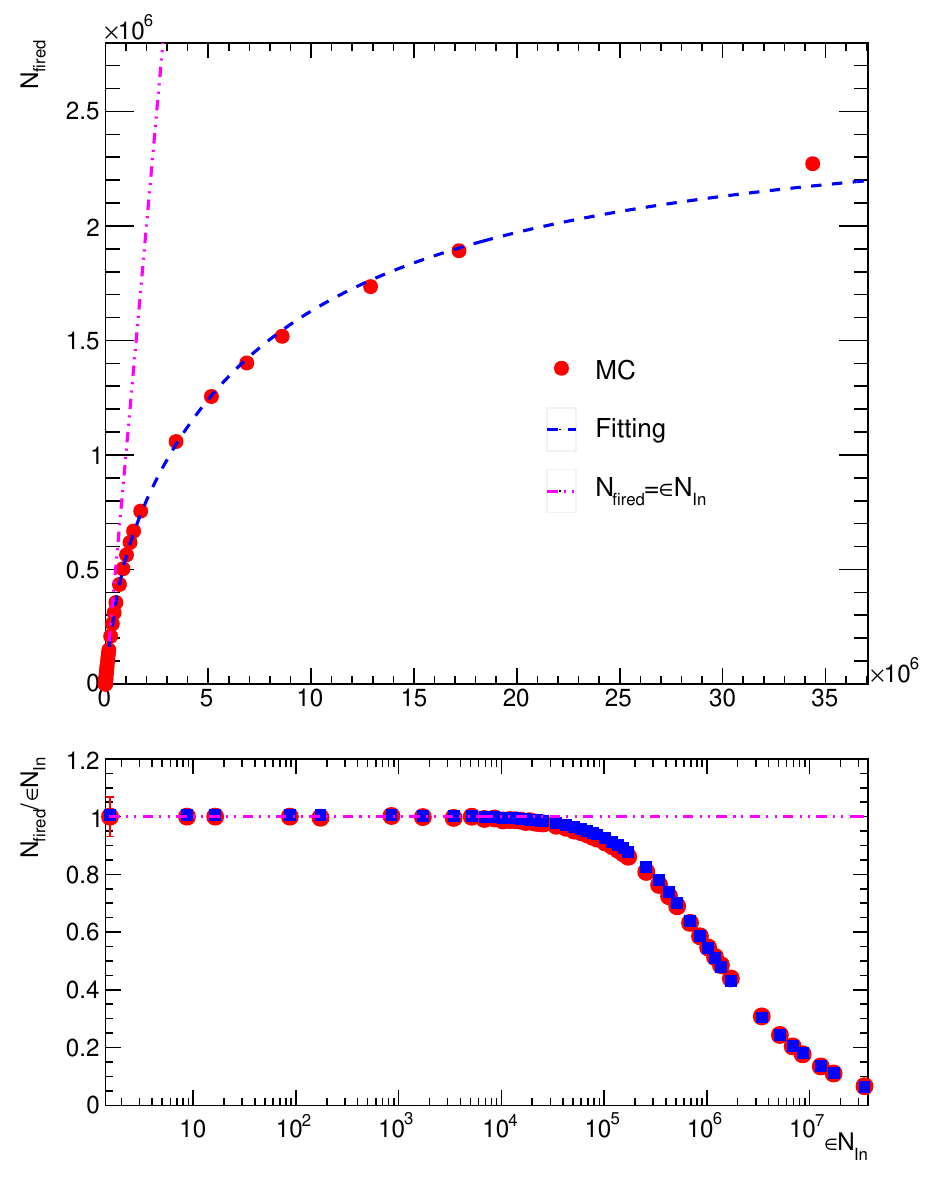}} 
\subfigure[]{
\includegraphics[width=0.32\textwidth]{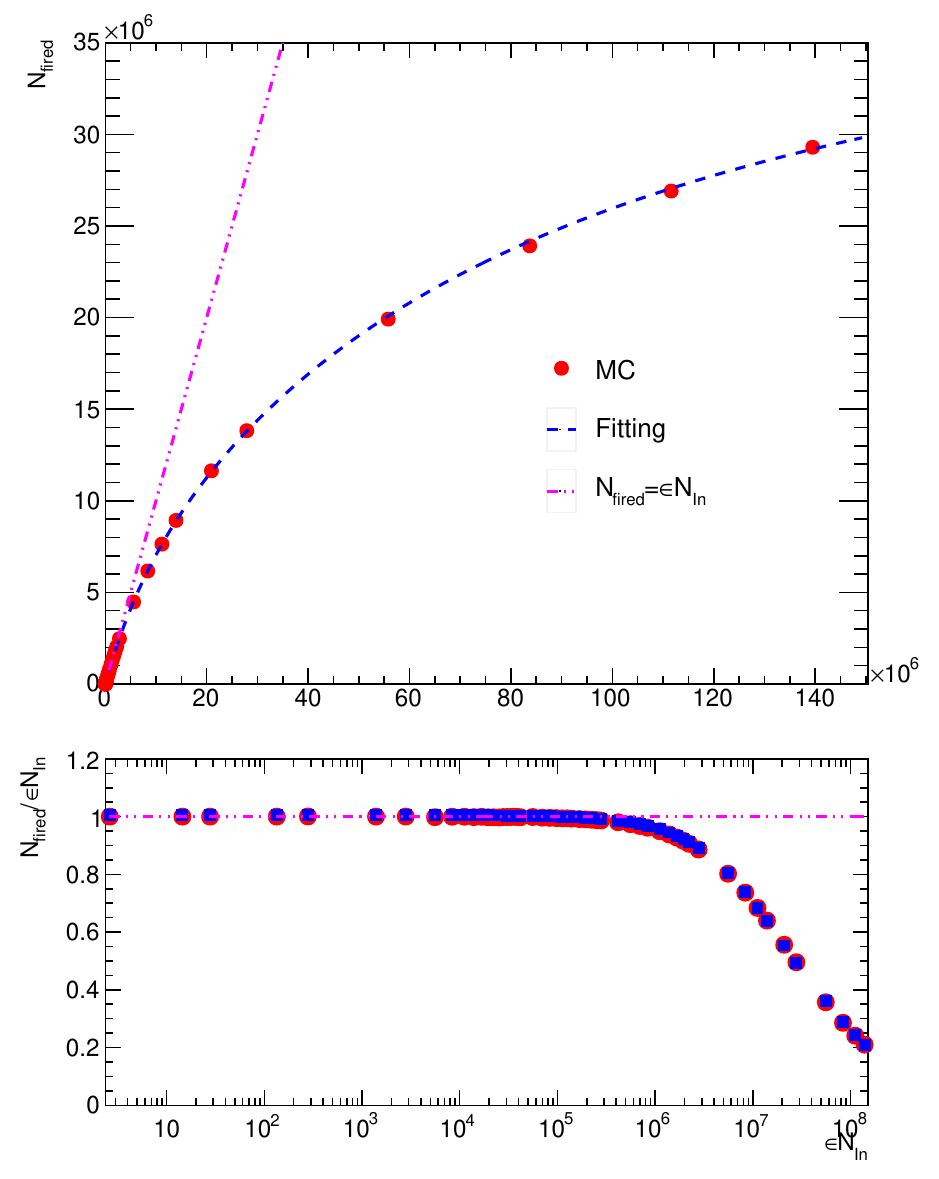}}
\caption{\label{fig:SiPMCurve_Rec}~Simulated responses of SiPMs to BGO scintillation light: (a) HAMAMATSU S13360-6025PE (b) HAMAMATSU S14160-3010PS (c) NDL EQR06 11-3030D-S. In the upper figures, the red points represent the simulated results, the blue dashed line corresponds to fitting functions from \cite{kotera2016}, and the purple dashed line represents a slope of one. In the lower figures, the blue and red points represent the ratio of actual photon counts to effective photon counts as measured by the SiPM from the fitting function and simulation, respectively, with the purple dashed line equal to one.}
\end{figure*}

The results of the simulation for the three types of SiPMs, accounting for the multi-firing effect on pixels, are shown in Figure~\ref{fig:SiPMCurve_Rec}. The blue dashed lines in the figures represent the fitting function as described in \cite{kotera2016}. When measuring the scintillation light from a BGO crystal bar, the SiPM response exhibits a significantly broader linear region. Compared to the laser simulation results in Figure~\ref{fig:SiPM_noRec}, the onset of nonlinear behavior is delayed by approximately an order of magnitude, primarily due to the long decay time of the BGO scintillation light. Table~\ref{tab:SiPMSatPoint_BGO} lists the effective photon count corresponding to 5\% nonlinearity for the three SiPMs.

\begin{table}[ht]
\centering
\caption{\label{tab:SiPMSatPoint_BGO} Effective photon counts at the deviation point of 5\% nonlinearity, measuring scintillation photon by BGO crystal bar.}
    \begin{tabular}{ccc}
        \toprule
        \makecell[c]{13360-6025PE} &\makecell[c]{S14160-3010PS}  &\makecell[c]{EQR06 11-3030D-S} \\
        \midrule
        19592 & 53747 & 1106210 \\
        \bottomrule
    \end{tabular}
\end{table}

Considering that the dynamic range of p.e. for a single channel in the CEPC crystal calorimeter can reach 350,000, the selected SiPM should ideally remain linear within this range. The results indicate that the SiPM with a total of 244,719 pixels and a 6-micron pixel size exhibits almost absolute linearity when detecting 350,000 scintillation photons from a BGO crystal bar, thereby meeting our dynamic range requirements. Although the device tests showed unexpected results, these findings demonstrate that such an SiPM configuration can fulfill the required performance criteria.

Conversely, the SiPM with a 10-micron pixel size and a total of 89,984 pixels shows approximately 20\% nonlinearity, leading to nonlinear measurements for high-energy events. However, the simulated response provided can be used to correct for SiPM saturation effects, offering a method to measure large signals with SiPMs having a smaller pixel count.

This approach can be applied not only to study the response of these three types of SiPMs to BGO crystal bars but also to investigate the response of different SiPMs to various scintillation crystals. This method allows for a comprehensive assessment of SiPM nonlinearity and the implementation of necessary corrections.

\section{Conclusions}
In this study, we performed a comprehensive evaluation of the intrinsic dynamic range of Silicon Photomultipliers (SiPMs) with varying pixel pitches of 6 $\mu$m, 10 $\mu$m, and 25 $\mu$m, each with pixel counts exceeding 50,000. Our experimental results revealed that SiPMs with 25 $\mu$m and 10 $\mu$m pixel pitches, corresponding to 57,600 and 89,984 pixels respectively, exhibited saturation levels that were slightly below their nominal pixel counts. In contrast, the SiPM with a 6 $\mu$m pixel pitch and 244,719 pixels showed a saturation point at approximately half of its nominal pixel count, which is not in expectation and need more investigations to unsdertand.

To gain deeper insights into these observations, we developed a detailed toy Monte Carlo simulation incorporating key attributes of SiPMs such as pixel density, photon detection efficiency (PDE), fill factor, avalanche triggering probability, crosstalk density, and the characteristics of BGO crystal scintillators. The simulation results for SiPMs' response to laser light demonstrated a consistent trend with the experimental data, showcasing a similar pattern of nonlinearity. For the SiPMs' response to BGO scintillation light, the simulations, which accounted for the multi-firing effect of pixels, indicated significantly broader linear region compared to the laser simulations. This discrepancy arises due to the longer decay time of BGO scintillation light and the necessity of considering the multi-firing effect of SiPM pixels.

Furthermore, our findings demonstrate that a system combining a BGO crystal bar of dimensions 40 $\times$ 40 $\times$ 1 cm$^3$ with SiPM featuring 244,719 pixels and 6 $\mu$m pixel size achieve approximately 100\% linearity to 350,000 scintillation photons, thus meeting the stringent dynamic range requirements for single channels in the CEPC crystal calorimeter. Conversely, the SiPM with 10 $\mu$m pixel size and 89,984 pixels exhibited approximately 20\% nonlinearity under similar conditions. However, the response curves provided in our study can be used to correct for SiPM saturation effects, making it feasible to measure high-energy events even with SiPMs having a limited number of pixels.

This study underscores the importance of considering pixel recovery effects and provides a robust framework for evaluating and optimizing SiPM performance in various high-energy physics applications. The methodologies and insights presented here can be extended to assess the nonlinearity and dynamic range of different SiPMs across various scintillation materials, thereby aiding in the design and development of more efficient and accurate photodetection systems.

\section*{Acknowledgments}
This work was supported by National Natural Science Foundation of China (Grant No.: 12150006), National Key R\&D Program of China (Grant No.: 2023YFA1606904 and 2023YFA1606900), and Shanghai Pilot Program for Basic Research—Shanghai Jiao Tong University (Grant No.: 21TQ1400209).

\bibliographystyle{unsrt}
\bibliography{main}  

\end{document}